\documentclass[%
 aip,
 amsmath,amssymb,
reprint, 
]{revtex4-1}

\usepackage{graphicx}
\usepackage{dcolumn}
\usepackage{bm}

\usepackage[utf8]{inputenc}
\usepackage[T1]{fontenc}
\usepackage{mathptmx}
\usepackage{etoolbox}

\makeatletter
\def\@email#1#2{%
 \endgroup
 \patchcmd{\titleblock@produce}
  {\frontmatter@RRAPformat}
  {\frontmatter@RRAPformat{\produce@RRAP{*#1\href{mailto:#2}{#2}}}\frontmatter@RRAPformat}
  {}{}
}%
\makeatother
\begin{document}

\preprint{AIP/123-QED}

\title{A 2D quantum dot array in planar $^{28}$Si/SiGe}
\author{F.K. Unseld}
\author{M. Meyer}
\affiliation{These authors contributed equally}
\affiliation{QuTech and Kavli Institute of Nanoscience, Delft University of Technology, Lorentzweg 1,2628 CJ Delft, The Netherlands}
\author{M.T. M\k{a}dzik}
\author{F. Borsoi}
\author{S.L. de Snoo}
\affiliation{QuTech and Kavli Institute of Nanoscience, Delft University of Technology, Lorentzweg 1,2628 CJ Delft, The Netherlands}

\author{S.V. Amitonov}
\affiliation{QuTech and Kavli Institute of Nanoscience, Delft University of Technology, Lorentzweg 1,2628 CJ Delft, The Netherlands}
\affiliation{QuTech and Netherlands Organization for Applied Scientific Research (TNO), Stieltjesweg 1, 2628 CK Delft, Netherlands}

\author{A. Sammak}
\affiliation{QuTech and Netherlands Organization for Applied Scientific Research (TNO), Stieltjesweg 1, 2628 CK Delft, Netherlands}

\author{G. Scappucci}
\author{M. Veldhorst}
\author{L.M.K. Vandersypen}
\affiliation{QuTech and Kavli Institute of Nanoscience, Delft University of Technology, Lorentzweg 1,2628 CJ Delft, The Netherlands}

\date{\today}

\begin{abstract}
Semiconductor spin qubits have gained increasing attention as a possible platform to host a fault-tolerant quantum computer. 
First demonstrations of spin qubit arrays have been shown in a wide variety of semiconductor materials. The highest performance for spin qubit logic has been realized in silicon, but scaling silicon quantum dot arrays in two dimensions has proven to be challenging. By taking advantage of high-quality heterostructures and carefully designed gate patterns, we are able to form a tunnel coupled 2 $\times$ 2 quantum dot array in a $^{28}$Si/SiGe heterostructure. We are able to load a single electron in all four quantum dots, thus reaching the (1,1,1,1) charge state. Furthermore we characterise and control the tunnel coupling between all pairs of dots by measuring polarisation lines over a wide range of barrier gate voltages. Tunnel couplings can be tuned from about $30~\rm \mu  eV$ up to approximately $400~\rm \mu eV$. These experiments provide a first step toward the operation of spin qubits in $^{28}$Si/SiGe quantum dots in two dimensions.
\end{abstract}

\maketitle
Since the original proposal for quantum computation with semiconductor quantum dots~\cite{Loss1998}, remarkable developments have been made. Quantum dot qubits are small in size, compatible with semiconductor manufacturing, and can be be operated with single-qubit gate fidelities and two-qubit gate fidelities above $99.9~\%$~\cite{Yoneda2017} and $99~\%$~\cite{Xue2022,Noiri2022,Mills2022} respectively. 

The implementation of two-dimensional qubit arrays will allow this technology platform to fully utilize its advantages. 
In GaAs heterostructures 2x2 and 3x3 quantum dot arrays have already been demonstrated~\cite{Dehollain2020,Mortemousque_2021,Fedele_2021}. However, hyperfine interaction leads to short dephasing times, preventing high-fidelity operation of qubit arrays. In contrast, group IV materials benefit from nuclear spin-free isotopes, such that quantum coherence can be maintained over much longer times~\cite{Stano2022}. 

In recent years, hole quantum dots in Ge/SiGe heterostructures progressed from a single quantum dot to a $4\times4$ quantum dot array with shared gate control \cite{Hendrickx2021,van_Riggelen_2022,Borsoi2022}. Parallel to that also silicon based devices have been pushed towards 2D arrays. Using quantum dots confined in the corners of silicon nanowires, several  $2\times\rm{N}$ quantum dot arrays have been demonstrated, albeit not simultaneously at the single-electron occupancy~\cite{Gilbert_2020, Chanrion_2020}. Furthermore, these devices did not contain separate gates for independent control of the tunnel barriers between neighbouring dots. This limits the controllability for quantum simulations and prevents sweet-spot operation~\cite{Martins2016,Reed2016} of exchange-based quantum gates.  

In this work, we present a 2D quantum dot array in gated planar $^{28}$Si/SiGe with barrier gates to control inter-dot tunnel couplings. Four quantum dots in a $2\times2$ configuration are formed with occupations controlled down to the last electron. Furthermore, all inter-dot tunnel couplings are characterized as a function of all barrier gate voltages. We demonstrate adequate tunnel coupling control and provide suggestions for future scalable gate designs.

The $2\times2$~quantum dot array investigated in this work is fabricated on a $^{28}$Si/Si$_{70}$Ge$_{30}$ heterostructure (see supplementary material). Figure~\ref{fig:Fig1}a shows a false-coloured scanning electron micrograph (SEM) image of a nominally identical device, highlighting the three gate layers of the multi-layer gate stack\cite{Lawrie2020}.
The screening gates in the first layer (purple) define an active area, reduce the formation of spurious dots and prohibit accumulation of a two-dimensional electron gas (2DEG) in the gate fan-out region. The second layer (yellow) consists of plunger (P) and accumulation gates. Barrier gates (B) are fabricated in the third layer (red). On top of the gate stack sits a micro magnet. The SEM image in Figure~\ref{fig:Fig1}a is taken before its deposition to highlight the quantum dot gate pattern. 

The gate stack defines four quantum dots in a $2\times2$~grid (labeled clockwise 1-4) and two single-electron transistors (SETs) (S1 and S2). The two SETs serve as charge detectors and act as electron reservoirs for the quantum dots in the $2\times2$~array. Off-chip NbTiN inductors connected to the SET reservoirs and parasitic capacitances form a tank circuit that enables radio frequency (RF) reflectrometry readout, allowing for fast and accurate detection of the charge occupation of all four quantum dots. 

\begin{figure*}[htb]
\centering
\includegraphics[width=0.98\textwidth]{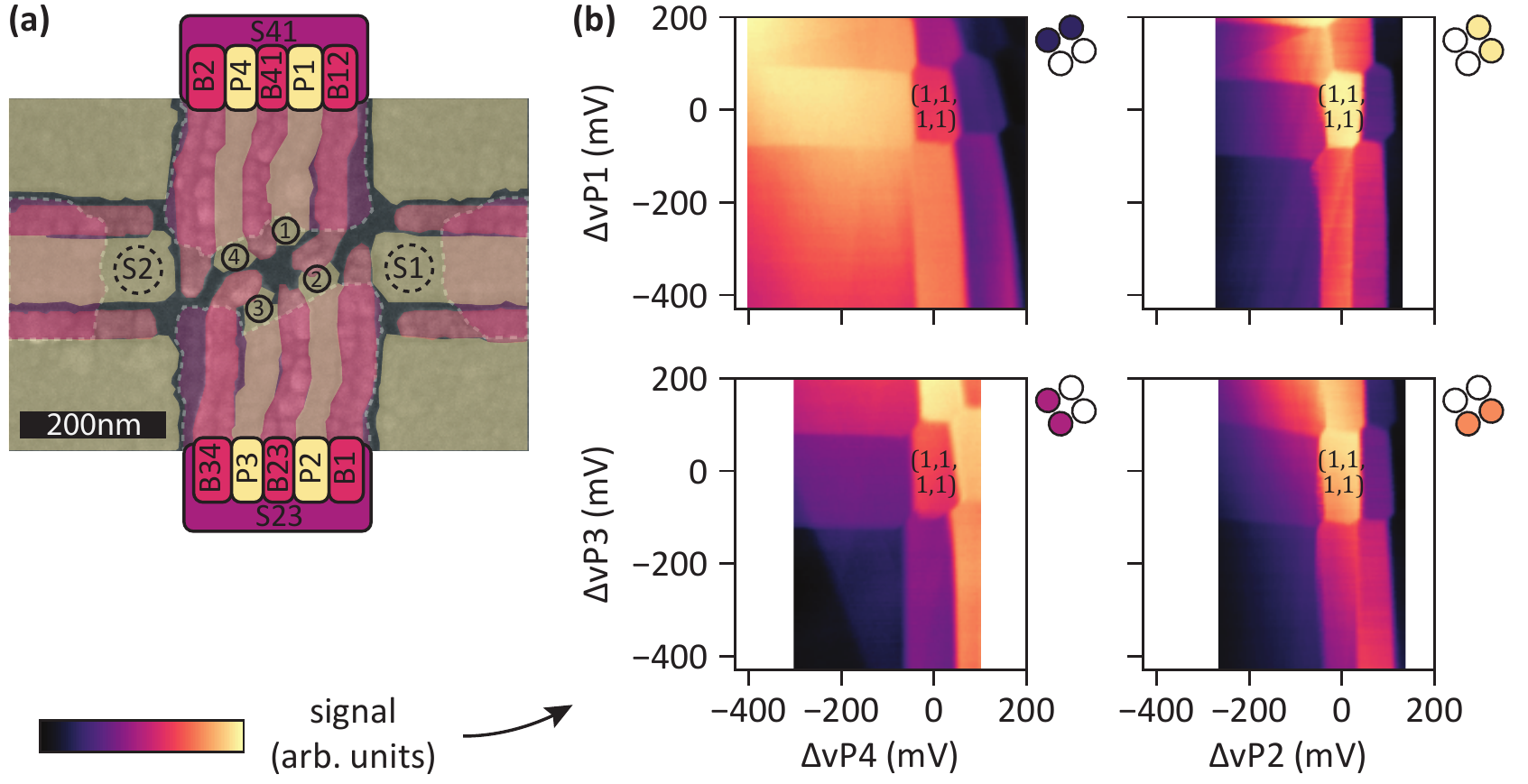}
\caption{\textbf{a)} False colored SEM image of a nominally identical device to the one used in the measurements. The four quantum dots in the center are labeled clockwise 1-4 with one sensor on each side marked as S1 and S2. Dashed lines mark the boundaries of the screening gates in the first gate layer.  \textbf{b)} Charge stability diagrams of nearest-neighbour quantum dots. Colored circles indicate the quantum dots of the swept virtual plunger gates while the quantum dots corresponding to the white circles remained with one electron each. All four scans thus show the (1,1,1,1) charge state where four quantum dots are occupied with one electron each. At $ \Delta v{\rm P}_{ i} = 0$ mV, the corresponding physical voltages on the gates are set to 2566~mV, 1831~mV, 3173~mV, 2487~mV for plungers 1-4, respectively. }
\label{fig:Fig1}
\end{figure*}

During the device tune-up, we measure the cross-capacitive coupling of all gates to all dots and virtualise them as described in \cite{Volk2019} with vP$_{i}$ (vB$_{ij}$) denoting the virtualised plunger (barrier) gates. The chosen virtual gates compensate the cross-talk onto all dot potentials and maintain the operation point of the charge sensors. The cross-capacitive coupling matrix $\mathbb{M}$, translating the real gate space to the virtual gate space via $\vec{V}^{\rm virt} = \mathbb{M} \vec{V}^{\rm real}$, is provided in the supplementary material. 

To show control over the charge occupation of the entire $2\times2$~array, we measure four charge stability diagrams as depicted in Figure~\ref{fig:Fig1}b. We acquire this data by sweeping the voltages on adjacent virtual plungers gates vP$_{i}$ and vP$_{(i~\text{mod}~4) +1}$ while monitoring the response of the charge sensors. The colored circles in the top right corner of each charge stability diagram indicate the position of the quantum dots corresponding to the swept plunger gates.

A honeycomb pattern characteristic of double-dot behavior is observed for all four plunger pairs. We identify the first electron in the four quantum dots by the absence of any more charge transitions in the lower left corner. Thus we can controllably access the $(N_{1},N_{2},N_{3},N_{4}) = (1,1,1,1)$ charge state, where $N_{i}$ denotes the charge occupation of quantum dot $\rm QD_{i}$, and isolate a single spin per quantum dot. The honeycomb patterns in Figure~\ref{fig:Fig1}b also show that all four quantum dots are capacitively coupled to each other. 

We note that there are apparent differences in the separation between the consecutive charge transition lines as well as in the slopes of successive charge transition lines. These could be caused by inherent differences and gate-voltage dependent variations in size, position or lever arm of the four intended quantum dots. Alternatively, they might be the charging signature of additional quantum dots in the close vicinity. While we cannot fully rule out the presence of such stray dots at higher occupations, we can reliably reach the (1,1,1,1) charge state in the $2\times 2$ configuration of the array. 

Next to the expected charge transitions, we observe additional diagonal features in Figure~\ref{fig:Fig1}b, which we associate with spurious defects in our system. These defects capacitively couple to the charge sensor but there is no indication of capacitive interaction with the four intentional quantum dots of the $2\times2$~array. 

Besides a well-defined charge state, controlled inter-dot tunnel couplings are essential for the implementation of robust exchange-based quantum gates or the execution of analog quantum simulation. Therefore, we probe the system evolution as a function of the voltage applied to the virtual barrier gates ${\rm vB}_{ij}$ located between the plunger gates of quantum dot $\rm QD_{i}$ and $\rm QD_{j}$ with $j=(i~\text{mod}~4)+1$. The tunnel coupling diagonally between $\rm QD_{1}$ and $\rm QD_{3}$ and anti-diagonally between $\rm QD_{2}$ and $\rm QD_{4}$ has no dedicated barrier gate and thus is not independently controllable. The influence of other barrier gates on the (anti-)diagonal tunnel coupling is presented below. 

\begin{figure*}[htb]
\centering
\includegraphics[width=0.98\textwidth]{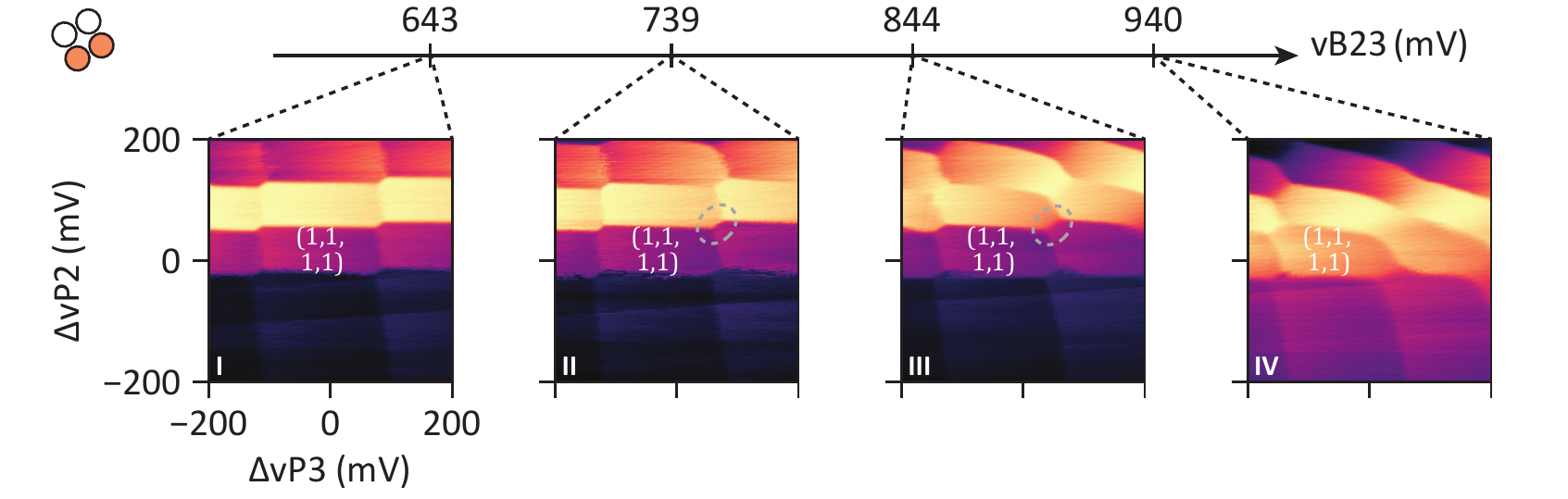}
\caption{Response of the charge stability diagram of $\rm QD_{\rm 2}$ and $\rm QD_{\rm 3}$ to changes of virtual barrier voltage $ {\rm vB}_{23}$, as indicated by the arrow on top. From panel I to IV, we observe a gradual increase in the both the capacitive and tunnel coupling between the two dots. Similar data were taken for all other nearest-neighbouring pairs and are displayed in the supplementary material. }
\label{fig:Fig2}
\end{figure*}

Figure \ref{fig:Fig2} shows the evolution of the charge stability diagram of $\rm QD_{2}$ and $\rm QD_{3}$ while changing the virtual barrier gate voltage ${\rm vB}_{23}$. The sequence of panels allows us to qualitatively asses the influence of the barrier on the capacitive coupling and tunnel coupling between the involved quantum dots. From panel I through IV, we observe that the separation between the  triple points increases, which indicates an increase in the capacitive coupling between the dots. In addition, we observe that the interdot charge transition is increasingly blurred (see the circled transitions) and the boundaries of the charge stability diagram are increasingly rounded. Both are indicative of an increased interdot tunnel coupling. In panel IV, for transition lines with $N_2+ N_3 \geq 4$ the rounding is so strong that the quantum dots have mostly merged into a single large dot.

\begin{figure*}[htb]
\centering
\includegraphics[width=0.98\textwidth]{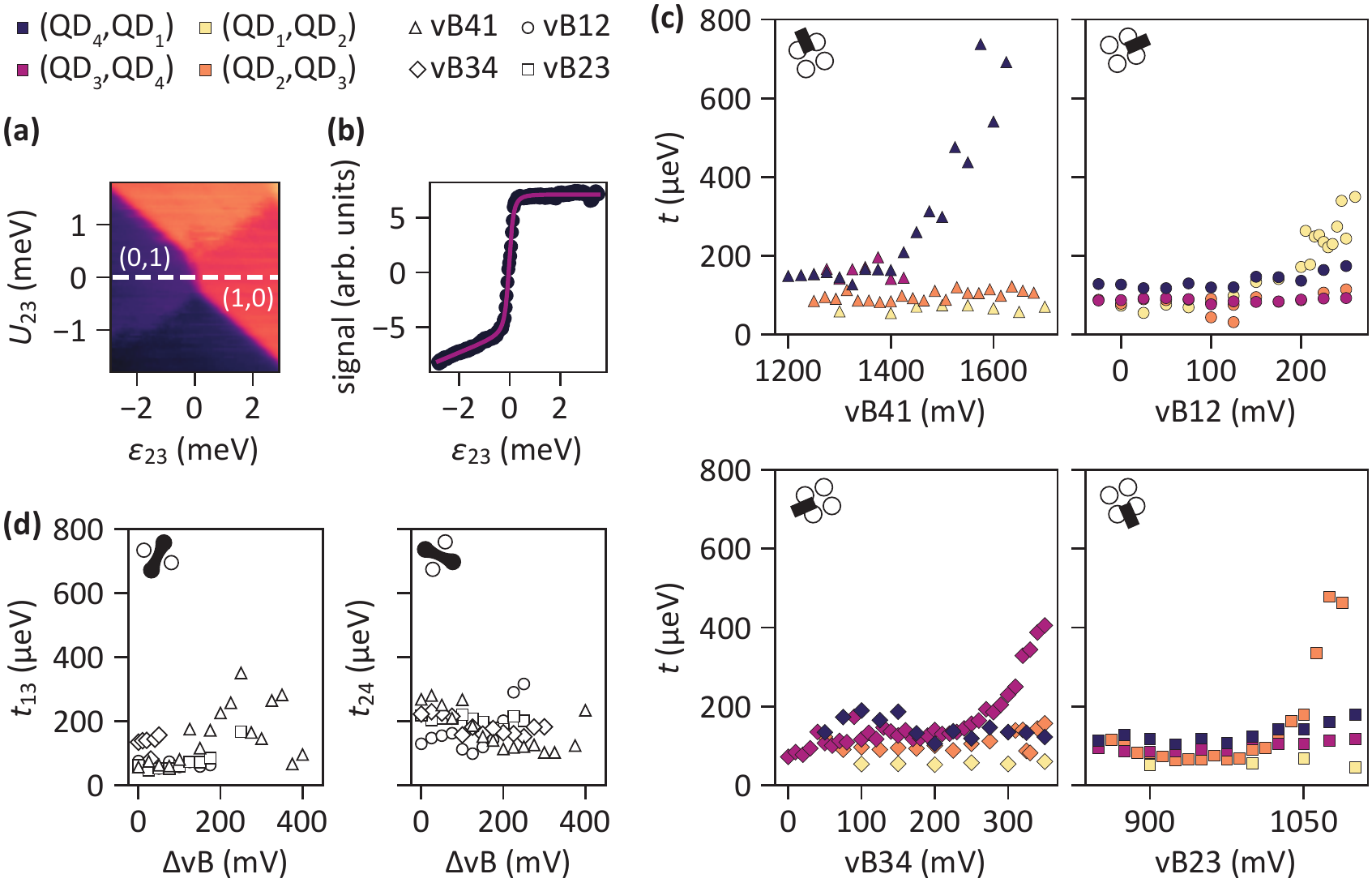}
\caption{\textbf{a)} Exemplary charge stability diagram around the (1,0) to (0,1) transition for $\rm QD_{2}$ and $\rm QD_{3}$ as a function of the interdot detuning $\epsilon_{23}$ and $U_{23}/\alpha_{U23} = \text{vP}_{2}+\text{vP}_{3}$. The dashed line indicates the detuning axis used to measure polarisation lines. \textbf{b)} Example of a measured (dots) and fitted (solid line) polarisation line for $\rm QD_{2}$ and $\rm QD_{3}$. \textbf{c)} Dependence of the tunnel couplings extracted from polarization lines between neighbouring quantum dots on each of the four virtual barrier gate voltages. The plots are ordered to follow the physical position of the barrier gate e.g. barrier $\rm vB_{41}$ situated in the top left corner of the quantum dot array is depicted in the top left plot. The legend for symbols and colors is shown above panels a) and b). We note that the dc voltages of barriers $\rm vB_{12}$ ($\rm vB_{23}$) and $\rm vB_{34}$ ($\rm vB_{41}$) are of comparable values, which is consistent with the symmetries of the gate pattern. Between scans we adjusted gate voltages of uninvolved gates to retain a high visibility. These adjustments were done in such a way that all uninvolved barrier gates remained in the small (residual) tunnel coupling regime.  On several occasions, insufficient contrast between the (1,0) and (0,1) charge states limited the data we were able to reliably fit. These data points are thus not available. \textbf{d)} Diagonal tunnel coupling and anti-diagonal tunnel coupling as a function of all four virtual barrier gate voltages. 
The physical gate voltages used at $\Delta\rm vB =0$ mV varies between data sets, as the voltages were slightly adjusted. As in c) uninvolved barrier gates remained in the residual tunnel coupling regime. Note also that in c) and d) the charge states vary between scans, depending on which dots each polarization line connects. 
}
\label{fig:Fig3}
\end{figure*}

To quantitatively determine the effect of the barrier voltage on the tunnel coupling, we measure polarisation lines along the detuning axis $\epsilon_{ij}/\alpha_{\epsilon ij} =\text{vP}_{i}-\text{vP}_{j}$, with $\alpha_{\epsilon ij}$ denoting the lever arm, across the  $(N_i, N_j) = (1,0)$ to $(0,1)$ interdot transition, as shown in Figure~\ref{fig:Fig3}a.  Scanning along this detuning axis moves the electron from dot 2 to dot 3, resulting in a step response in the sensor signal as seen in figure \ref{fig:Fig3}b. 
This step response is broadened by both the electron temperature $T_e \leq 78.5\pm 2.2~\rm~mK$ and the tunnel coupling $t$, and can be fitted using $S_{\rm Sig} = \frac{\epsilon}{\Omega}\tanh{\frac{\Omega}{2{k_b T_e}}}$ with $\Omega = \sqrt{\epsilon^2+4 t^2}$ and $\epsilon$ the detuning between the two quantum dots \cite{DiCarlo2004}. Additional slopes and offsets of the sensor signal caused by imperfect virtualisation or drifts are taken into account in the used fitting procedure~\cite{vanDiepen2018}.

We systematically extract the dependency of the inter-dot tunnel couplings $t_{n,m}$ between all dot pairs $({\rm QD}_{n},{\rm QD}_{m})$ with respect to all barrier voltages ${\rm vB}_{ij}$. Figure \ref{fig:Fig3}c shows the resulting tunnel couplings $t_{n,m}$ grouped by barrier gates ${\rm vB}_{ij}$. We observe that changing the barrier voltage ${\rm vB}_{ij}$ affects only the corresponding tunnel couplings $t_{ij}$ significantly, while keeping the other tunnel couplings largely constant. Note that the virtual gate matrix compensates for cross-talk of the barrier gates onto all dot potentials, but does not account for possible cross-talk on the tunnel couplings. 

We furthermore find that below a given voltage (which is different for each ${\rm vB}_{ij}$), the influence of the barrier gate voltage on the corresponding tunnel coupling vanishes and a residual tunnel coupling remains. Across all four neighbouring dot pairs, the residual tunnel coupling is in the range between $30~\rm \mu eV$ and $200~\rm\mu eV$. 

We extend this characterisation to the (anti-)diagonal tunnel couplings. Figure \ref{fig:Fig3}d  presents the influence of the four barrier gates on the diagonal and anti-diagonal tunnel coupling respectively.  
While the anti-diagonal tunnel coupling $t_{2,4}$ is elevated and can be modulated using ${\rm vB}_{12}$ in particular, the diagonal tunnel coupling $t_{1,3}$ is not systematically influenced by any barrier gate and remains in many cases lower than all other tunnel couplings, albeit far from zero.

We demonstrated the first 2D quantum dot array in a planar silicon technology and operated the four quantum dots in the single electron regime, consistently achieving the $(1,1,1,1)$ charge state. 
Furthermore, the barrier gates allow us to independently control the interdot tunnel couplings. However, the residual tunnel couplings observed in this sample are higher than the typical tunnel coupling of 1-10 $\mu$eV used in spin qubit experiments~\cite{Knapp2016}. Presumably the close proximity of the screening gates to the center of the plunger gates compresses the quantum dots towards the center of the $2\times 2$ array and hence towards each other, leading to rather large tunnel couplings. Furthermore, we see in Fig.~\ref{fig:Fig3} that at low tunnel coupling values, the tunnel coupling barely responds to the barrier gate voltages anymore. 
The compressed position of the quantum dots in the center region enhances also the diagonal coupling between them. While analog quantum simulation and quantum computation can benefit from diagonal tunnel coupling, the lack of dedicated control over magnitude and directionality i.e. diagonal versus anti-diagonal, also poses limitations. Suppressing any diagonal coupling with a center gate as demonstrated in a GaAs $2 \times 2$ array could be a suitable way to circumvent this issue~\cite{Mukhopadhyay2018}. 

The encountered challenges help to identify possible improvements in the design of planar $2\times 2$ $^{28}$Si/SiGe quantum dot arrays. Specifically, moving the screening gates away from the center of the array is expected to yield lower tunnel couplings, as the electrons are not squeezed towards each other as much. The experiments also offer relevant learnings for scaling to larger arrays. For instance, changing the device architecture from a square array to a triangular array will alleviate the issues regarding undesired diagonal tunnel couplings. Furthermore, in order to maintain control of  individual tunnel couplings, either more sophisticated patterning techniques must be applied to route gates to the inside of a larger array~\cite{Ha2021}, or crossbar addressing must be employed~\cite{Veldhorst2017, Li2018, Borsoi2022}. In both cases, the observations made for the present device provide guidance for suitable plunger and barrier gate pitches and dimensions. 

\begin{acknowledgments}
We acknowledge funding by Intel Corporation and fruitful discussions provided by everyone in the Veldhorst Lab and Vandersypen Lab. 
\end{acknowledgments}

\section*{Conflict of Interest}
The authors have no conflicts to disclose.
\section*{Author Contributions}

\textbf{F.K. Unseld}: Conceptualization (lead), Data curation (equal), Formal analysis (equal), Investigation (lead), Methodology (equal), Resources (equal), Validation (equal), Visualization (equal), Writing – original draft (lead), Writing – review \& editing (lead).
\textbf{M. Meyer}: Conceptualization (lead), Data curation (lead), Formal analysis (lead), Investigation (lead), Methodology (equal), Validation (equal), Visualization (lead), Writing – original draft (equal), Writing – review \& editing (equal).
\textbf{M.T. M\k{a}dzik}: Formal analysis (supporting), Investigation (supporting), Supervision (supporting), Validation (supporting), Writing – review \& editing (supporting).
\textbf{F. Borsoi}: Formal analysis (supporting), Investigation (supporting), Methodology (supporting),  Software (equal), Validation (supporting), Writing – review \& editing (supporting).
\textbf{S.L. de Snoo}: Software (lead), Writing – review \& editing (supporting).
\textbf{S.V. Amitonov}: Resources (equal), Writing – review \& editing (supporting).
\textbf{A. Sammak}: Resources (equal), Writing – review \& editing (supporting).
\textbf{G. Scappucci}: Resources (equal), Writing – review \& editing (supporting).
\textbf{M. Veldhorst}: Conceptualization (lead), Funding acquisition (lead), Project administration (lead), Supervision (lead), Validation (equal), Writing – review \& editing (equal).
\textbf{L.M.K. Vandersypen}: Conceptualization (lead), Funding acquisition (lead), Project administration (lead), Supervision (lead), Validation (equal), Writing – review \& editing (lead).

\section*{Data Availability Statement}
The data and analysis scripts that support the findings of this study are openly available in the Zenodo repository at http://doi.org/10.5281/zenodo.7957631, reference number \cite{Dataset}

\appendix

\section{Device fabrication and screening}
This device is fabricated on a $^{28}$Si/Si$_{70}$Ge$_{30}$ heterostructure. A $2.5~\rm \mu m$ strain relaxed Si$_{70}$Ge$_{30}$ buffer layer makes the foundation. On top of it the isotopically enriched $^{28}$Si quantum well is grown. It has a residual $^{29}$Si concentration of 0.08\% and was measured to be $9.0\pm0.5~\rm nm$ thick. Afterwards a $30~\rm nm$ thick Si$_{70}$Ge$_{30}$ spacer is grown to reduce strain relaxation in the quantum well and separate it from the gate dielectric. The heterostructure is finalized with a $1~\rm nm$ silicon cap~\cite{DegliEsposti2022}. The gate stack is separated from the heterostructure by $10~\rm nm$ Al$_2$O$_3$, formed by atomic-layer deposition (ALD) at 300 $^\circ$C. The three gate layers of the gate stack are made from Ti:Pd with thicknesses of 3:17, 3:27 and 3:27 nm and are patterned using electron beam lithography, electron beam evaporation and lift off. Each layer is electrically isolated from the previous layer by a $5~\rm nm$ Al$_2$O$_3$ dielectric grown by ALD. Above the three gate layers a micro magnet is fabricated from Ti:Co (5:200 nm). 

After fabrication every device was screened by measuring turn-on curves and testing for gate leakage or shorted gates at $4.2~\rm K$ with a dipstick in liquid helium. This high turnaround testing allows to verify the basic functionality of the device and quickly filter out defective devices. After verifying the functionality of all gates, NbTiN inductors are added to the ohmic contacts to enable RF-readout. To address electrostatic discharge concerns during rebonding, we screen the device a second time in liquid helium to verify that no damage has been done. Then we cool down the devices in a Bluefors LD400 dilution refrigerator to its base temperature of around $10~\rm mK$. 

\section{Setup}
The room-temperature control electronics to operate this device are separated into ac electronics in one rack and dc electronics in a second one. In the latter, several in-house built Serial Peripheral Interface (SPI) racks host $18\rm~bit$ Digital to Analog Converter (DAC) modules which provide the required dc voltages. Voltage dividers were used to apply an accurate source-drain bias when needed. The currents are measured with a Keithley 2000 Multimeter (placed in the ac rack) via an in-house developed transconductance amplifier. The dc rack is powered by batteries which are continuously charged via gyrators and filters. 

The ac rack comprises the host computer, a Keysight chassis (M9019A) and an additional dedicated RF SPI rack. The circuitry for RF reflectometroy measurements consists of two in-house built RF sources, a combiner (Mini-Circuits ZFSC-2-5-S+), a $15\rm~dB$ coupler (Mini-Circuits ZEDC-15-2B) at the mixing chamber stage, a cryo-amplifier at the 4K stage (Cosmic Microwave Technology Inc. CITLF3), a room temperature amplifier, two IQ-mixers and a Keysight digitizer (M3102A). The coaxial lines from 4 K to the mixing chamber flange are made from NbTiN to ensure a high signal quality and low thermal conductance. From 4 K to room temperature, SCuNi-CuNi cables are used. Discrete attenuators with a total attenuation of $23\rm~dB$ are distributed over the various temperature stages on the downward path. 

Next to the digitizer, several Keysight AWG modules (M3202A) are situated in the same chassis and connected via a PCIe connection to a host computer. SCuNi-CuNi $0.86\rm~mm$ coaxial cables are used from room temperature to the mixing chamber plate. Also on these lines discrete attenuators are mounted, with a total attenuation ranging from $12\rm~dB$ to $20\rm~dB$ (typically we equip the barrier gates with lower attenuation than the plunger gates). From the mixing chamber flange to the sample printed circuit board (PCB), hand-formable 0.086" coaxial cables were used to route both RF and AWG signals. Bias tees on the sample PCB combine the ac pulses and dc voltages. Ferrite cores and dc blocks were installed at room temperature to suppress $50\rm~Hz$ noise.   

\section{Virtual Gate Matrix}
\label{chpt:VGM}
\begin{figure}[htb]
\centering
\includegraphics[width=0.45\textwidth]{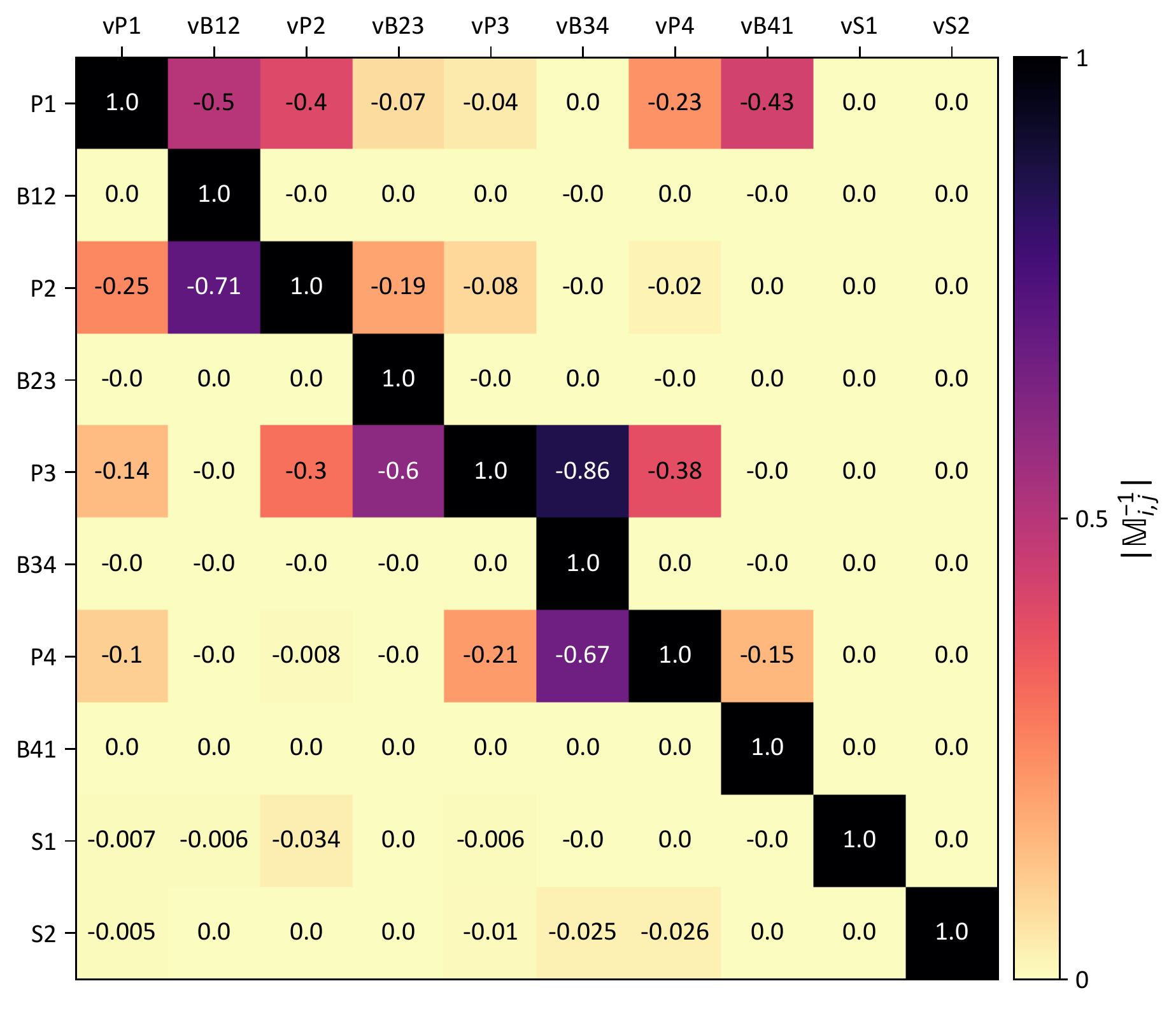}
\caption{Virtual gate matrix ($\mathbb{M}^{-1}$) during the recording of the charge stability diagrams of Fig.~\ref{fig:Fig1}}
\label{fig:VGM_Supp}
\end{figure}

The virtual gate matrix defining the virtual gates in the charge stability diagrams of Fig.~\ref{fig:Fig1} is displayed in Fig.~\ref{fig:VGM_Supp}. Note that we constantly adapted the virtual gate matrices throughout the measurements to improve the compensation of cross capacitive effects on the plunger gates of the quantum dots and SETs. 

\section{Qualitative Tunnel Coupling Control}
\label{chpt:QualTC} 
\begin{figure*}[htb]
\centering
\includegraphics[width=0.98\textwidth]{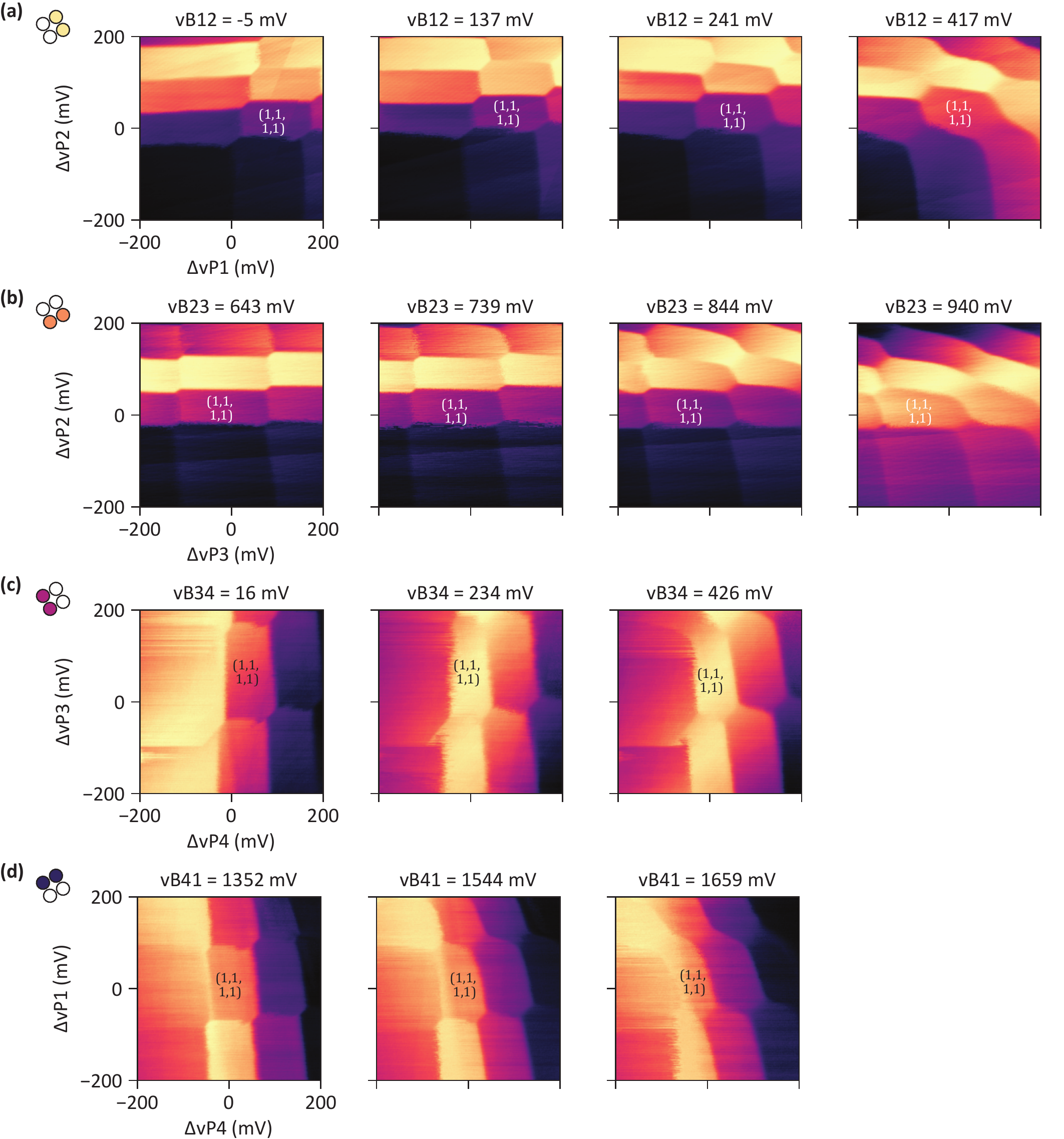}
\caption{Charge stability diagrams for all four quantum dot pairs for several values of the voltage on the barrier gate between the respective dots. Pictograms on the left indicate the quantum dot pair used in the respective row. To ensure comparability every scan is around the (1,1,1,1) regime as indicated.  }
\label{fig:QualTcControl_Supp}
\end{figure*}

Figure \ref{fig:QualTcControl_Supp} shows the influence of all barrier gates on the charge stability diagrams of their neighbouring quantum dots. Figure \ref{fig:QualTcControl_Supp}b is replicated from Figure \ref{fig:Fig2}. All barriers show similar influence on the charge stability diagram as discussed by way of example for barrier ${\rm vB}_{23}$ in the main text.

\section{Electron Temperature and Detuning Lever Arm Calculation}
\label{chpt:LA} 
\begin{figure*}[htb]
\centering
\includegraphics[width=0.7\textwidth]{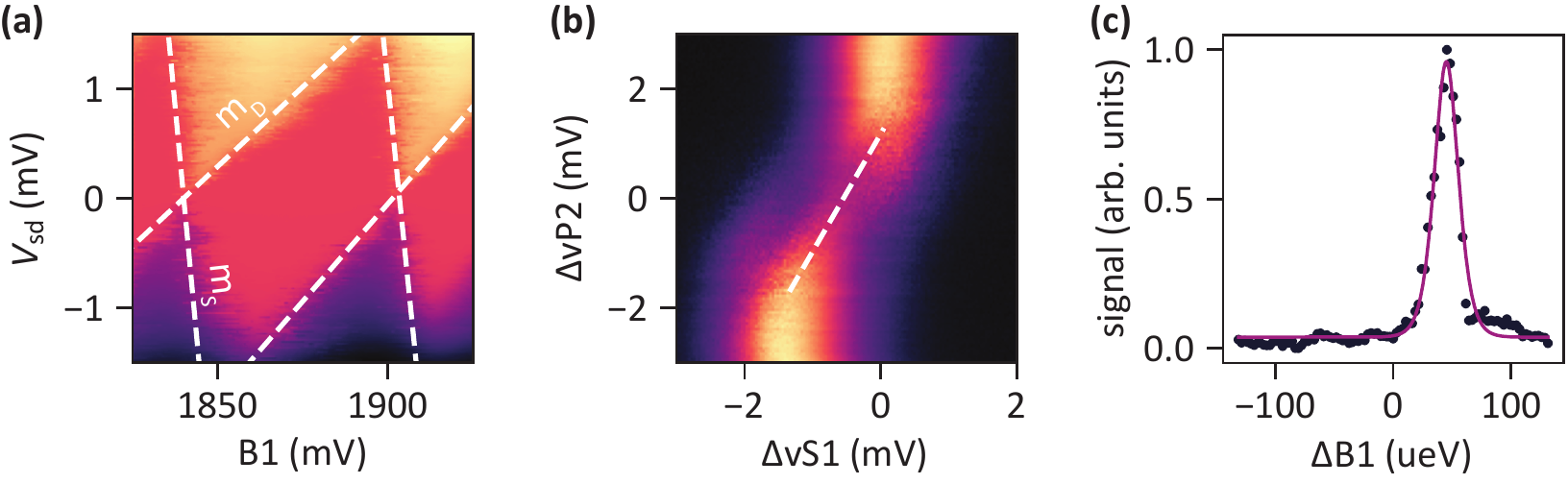}
\caption{\textbf{a)} Coulomb diamonds used to extract the lever arm of gate B1. White dashed lines indicate the used slopes. \textbf{b)} High resolution interdot transition of sensing dot S1 and $\rm QD_{2}$ showing the manual fit (dashed line) to determine the lever arm ratio between the respective plunger gates. \textbf{c)} Measured (dots) and fitted (solid line) Coulomb peak used to estimate the electron temperature. The Coulomb peak is the smallest measured peak at the effective zero bias voltage. }
\label{fig:Sup_detLA}
\end{figure*}

To estimate the electron temperature, we measure the thermal broadening of Coulomb peaks of SET S1 using the equations provided by~\cite{Petit2018}. We sweep gate $\rm B_{1}$ as it has a smaller lever arm compared to the SET plunger gate and therefore allows us to sweep across a Coulomb peak with a much finer resolution, improving the fit quality. Figure~\ref{fig:Sup_detLA}a shows the Coulomb diamonds that were used to calculate the lever arm of gate B1. We combine the slopes $m_S$ and $m_D$ of the Coulomb diamonds to compute the lever arm $\alpha_{B1}$ with $\alpha_{B1}=\lvert\frac{m_s m_D}{m_s-m_D}\rvert$. The horizontal trace at $V_{\rm SD} \approx -1150~\rm \mu V$ shown in Fig.~\ref{fig:Sup_detLA}c was used to upper bound the electron temperature to $\rm T_e \leq 80~\rm~mK$.

Furthermore we require the lever arms of the plunger gates to convert the detuning axis $\epsilon_{ij}$ from gate voltage to energy. Due to the high tunnel coupling, photon-assisted tunneling measurements with the available microwave source were unsuccessful. Instead we estimate the lever arms of the quantum dots via the slope of their interdot transitions. 

Using Coulomb diamond measurements (see Fig~\ref{fig:Sup_detLA}a) we extract the lever arm of gate $\rm B_{1}$. We convert the lever arm of gate $\rm vB_{1}$ to the lever arm of virtual plunger gate $\rm vS_{1}$ from the ratio of the SET Coulomb peak spacing. From here we can successively use the angle of the interdot transitions to calculate the lever arm ratio between all other gates. For example, Fig.~\ref{fig:Sup_detLA}b shows the manually fitted interdot transition used to calculate the lever arm ratio of plunger gate of $\rm vS_{1}$ and $\rm vP_{2}$.

\section*{References}
\bibliography{2DQuantumDotArrayInPlanarSiSiGe}

\begin{thebibliography}{28}%
\makeatletter
\providecommand \@ifxundefined [1]{%
 \@ifx{#1\undefined}
}%
\providecommand \@ifnum [1]{%
 \ifnum #1\expandafter \@firstoftwo
 \else \expandafter \@secondoftwo
 \fi
}%
\providecommand \@ifx [1]{%
 \ifx #1\expandafter \@firstoftwo
 \else \expandafter \@secondoftwo
 \fi
}%
\providecommand \natexlab [1]{#1}%
\providecommand \enquote  [1]{``#1''}%
\providecommand \bibnamefont  [1]{#1}%
\providecommand \bibfnamefont [1]{#1}%
\providecommand \citenamefont [1]{#1}%
\providecommand \href@noop [0]{\@secondoftwo}%
\providecommand \href [0]{\begingroup \@sanitize@url \@href}%
\providecommand \@href[1]{\@@startlink{#1}\@@href}%
\providecommand \@@href[1]{\endgroup#1\@@endlink}%
\providecommand \@sanitize@url [0]{\catcode `\\12\catcode `\$12\catcode
  `\&12\catcode `\#12\catcode `\^12\catcode `\_12\catcode `\%12\relax}%
\providecommand \@@startlink[1]{}%
\providecommand \@@endlink[0]{}%
\providecommand \url  [0]{\begingroup\@sanitize@url \@url }%
\providecommand \@url [1]{\endgroup\@href {#1}{\urlprefix }}%
\providecommand \urlprefix  [0]{URL }%
\providecommand \Eprint [0]{\href }%
\providecommand \doibase [0]{http://dx.doi.org/}%
\providecommand \selectlanguage [0]{\@gobble}%
\providecommand \bibinfo  [0]{\@secondoftwo}%
\providecommand \bibfield  [0]{\@secondoftwo}%
\providecommand \translation [1]{[#1]}%
\providecommand \BibitemOpen [0]{}%
\providecommand \bibitemStop [0]{}%
\providecommand \bibitemNoStop [0]{.\EOS\space}%
\providecommand \EOS [0]{\spacefactor3000\relax}%
\providecommand \BibitemShut  [1]{\csname bibitem#1\endcsname}%
\let\auto@bib@innerbib\@empty
\bibitem [{\citenamefont {Loss}\ and\ \citenamefont
  {DiVincenzo}(1998)}]{Loss1998}%
  \BibitemOpen
  \bibfield  {author} {\bibinfo {author} {\bibfnamefont {D.}~\bibnamefont
  {Loss}}\ and\ \bibinfo {author} {\bibfnamefont {D.~P.}\ \bibnamefont
  {DiVincenzo}},\ }\bibfield  {title} {\enquote {\bibinfo {title} {Quantum
  computation with quantum dots},}\ }\href {\doibase 10.1103/physreva.57.120}
  {\bibfield  {journal} {\bibinfo  {journal} {Physical Review A}\ }\textbf
  {\bibinfo {volume} {57}},\ \bibinfo {pages} {120--126} (\bibinfo {year}
  {1998})}\BibitemShut {NoStop}%
\bibitem [{\citenamefont {Yoneda}\ \emph {et~al.}(2017)\citenamefont {Yoneda},
  \citenamefont {Takeda}, \citenamefont {Otsuka}, \citenamefont {Nakajima},
  \citenamefont {Delbecq}, \citenamefont {Allison}, \citenamefont {Honda},
  \citenamefont {Kodera}, \citenamefont {Oda}, \citenamefont {Hoshi},
  \citenamefont {Usami}, \citenamefont {Itoh},\ and\ \citenamefont
  {Tarucha}}]{Yoneda2017}%
  \BibitemOpen
  \bibfield  {author} {\bibinfo {author} {\bibfnamefont {J.}~\bibnamefont
  {Yoneda}}, \bibinfo {author} {\bibfnamefont {K.}~\bibnamefont {Takeda}},
  \bibinfo {author} {\bibfnamefont {T.}~\bibnamefont {Otsuka}}, \bibinfo
  {author} {\bibfnamefont {T.}~\bibnamefont {Nakajima}}, \bibinfo {author}
  {\bibfnamefont {M.~R.}\ \bibnamefont {Delbecq}}, \bibinfo {author}
  {\bibfnamefont {G.}~\bibnamefont {Allison}}, \bibinfo {author} {\bibfnamefont
  {T.}~\bibnamefont {Honda}}, \bibinfo {author} {\bibfnamefont
  {T.}~\bibnamefont {Kodera}}, \bibinfo {author} {\bibfnamefont
  {S.}~\bibnamefont {Oda}}, \bibinfo {author} {\bibfnamefont {Y.}~\bibnamefont
  {Hoshi}}, \bibinfo {author} {\bibfnamefont {N.}~\bibnamefont {Usami}},
  \bibinfo {author} {\bibfnamefont {K.~M.}\ \bibnamefont {Itoh}}, \ and\
  \bibinfo {author} {\bibfnamefont {S.}~\bibnamefont {Tarucha}},\ }\bibfield
  {title} {\enquote {\bibinfo {title} {A quantum-dot spin qubit with coherence
  limited by charge noise and fidelity higher than 99.9{\%}},}\ }\href
  {\doibase 10.1038/s41565-017-0014-x} {\bibfield  {journal} {\bibinfo
  {journal} {Nature Nanotechnology}\ }\textbf {\bibinfo {volume} {13}},\
  \bibinfo {pages} {102--106} (\bibinfo {year} {2017})}\BibitemShut {NoStop}%
\bibitem [{\citenamefont {Xue}\ \emph {et~al.}(2022)\citenamefont {Xue},
  \citenamefont {Russ}, \citenamefont {Samkharadze}, \citenamefont {Undseth},
  \citenamefont {Sammak}, \citenamefont {Scappucci},\ and\ \citenamefont
  {Vandersypen}}]{Xue2022}%
  \BibitemOpen
  \bibfield  {author} {\bibinfo {author} {\bibfnamefont {X.}~\bibnamefont
  {Xue}}, \bibinfo {author} {\bibfnamefont {M.}~\bibnamefont {Russ}}, \bibinfo
  {author} {\bibfnamefont {N.}~\bibnamefont {Samkharadze}}, \bibinfo {author}
  {\bibfnamefont {B.}~\bibnamefont {Undseth}}, \bibinfo {author} {\bibfnamefont
  {A.}~\bibnamefont {Sammak}}, \bibinfo {author} {\bibfnamefont
  {G.}~\bibnamefont {Scappucci}}, \ and\ \bibinfo {author} {\bibfnamefont
  {L.~M.~K.}\ \bibnamefont {Vandersypen}},\ }\bibfield  {title} {\enquote
  {\bibinfo {title} {Quantum logic with spin qubits crossing the surface code
  threshold},}\ }\href {\doibase 10.1038/s41586-021-04273-w} {\bibfield
  {journal} {\bibinfo  {journal} {Nature}\ }\textbf {\bibinfo {volume} {601}},\
  \bibinfo {pages} {343--347} (\bibinfo {year} {2022})}\BibitemShut {NoStop}%
\bibitem [{\citenamefont {Noiri}\ \emph {et~al.}(2022)\citenamefont {Noiri},
  \citenamefont {Takeda}, \citenamefont {Nakajima}, \citenamefont {Kobayashi},
  \citenamefont {Sammak}, \citenamefont {Scappucci},\ and\ \citenamefont
  {Tarucha}}]{Noiri2022}%
  \BibitemOpen
  \bibfield  {author} {\bibinfo {author} {\bibfnamefont {A.}~\bibnamefont
  {Noiri}}, \bibinfo {author} {\bibfnamefont {K.}~\bibnamefont {Takeda}},
  \bibinfo {author} {\bibfnamefont {T.}~\bibnamefont {Nakajima}}, \bibinfo
  {author} {\bibfnamefont {T.}~\bibnamefont {Kobayashi}}, \bibinfo {author}
  {\bibfnamefont {A.}~\bibnamefont {Sammak}}, \bibinfo {author} {\bibfnamefont
  {G.}~\bibnamefont {Scappucci}}, \ and\ \bibinfo {author} {\bibfnamefont
  {S.}~\bibnamefont {Tarucha}},\ }\bibfield  {title} {\enquote {\bibinfo
  {title} {Fast universal quantum gate above the fault-tolerance threshold in
  silicon},}\ }\href {\doibase 10.1038/s41586-021-04182-y} {\bibfield
  {journal} {\bibinfo  {journal} {Nature}\ }\textbf {\bibinfo {volume} {601}},\
  \bibinfo {pages} {338--342} (\bibinfo {year} {2022})}\BibitemShut {NoStop}%
\bibitem [{\citenamefont {Mills}\ \emph {et~al.}(2022)\citenamefont {Mills},
  \citenamefont {Guinn}, \citenamefont {Gullans}, \citenamefont {Sigillito},
  \citenamefont {Feldman}, \citenamefont {Nielsen},\ and\ \citenamefont
  {Petta}}]{Mills2022}%
  \BibitemOpen
  \bibfield  {author} {\bibinfo {author} {\bibfnamefont {A.~R.}\ \bibnamefont
  {Mills}}, \bibinfo {author} {\bibfnamefont {C.~R.}\ \bibnamefont {Guinn}},
  \bibinfo {author} {\bibfnamefont {M.~J.}\ \bibnamefont {Gullans}}, \bibinfo
  {author} {\bibfnamefont {A.~J.}\ \bibnamefont {Sigillito}}, \bibinfo {author}
  {\bibfnamefont {M.~M.}\ \bibnamefont {Feldman}}, \bibinfo {author}
  {\bibfnamefont {E.}~\bibnamefont {Nielsen}}, \ and\ \bibinfo {author}
  {\bibfnamefont {J.~R.}\ \bibnamefont {Petta}},\ }\bibfield  {title} {\enquote
  {\bibinfo {title} {Two-qubit silicon quantum processor with operation
  fidelity exceeding 99{\%}},}\ }\href {\doibase 10.1126/sciadv.abn5130}
  {\bibfield  {journal} {\bibinfo  {journal} {Science Advances}\ }\textbf
  {\bibinfo {volume} {8}} (\bibinfo {year} {2022}),\
  10.1126/sciadv.abn5130}\BibitemShut {NoStop}%
\bibitem [{\citenamefont {Dehollain}\ \emph {et~al.}(2020)\citenamefont
  {Dehollain}, \citenamefont {Mukhopadhyay}, \citenamefont {Michal},
  \citenamefont {Wang}, \citenamefont {Wunsch}, \citenamefont {Reichl},
  \citenamefont {Wegscheider}, \citenamefont {Rudner}, \citenamefont {Demler},\
  and\ \citenamefont {Vandersypen}}]{Dehollain2020}%
  \BibitemOpen
  \bibfield  {author} {\bibinfo {author} {\bibfnamefont {J.~P.}\ \bibnamefont
  {Dehollain}}, \bibinfo {author} {\bibfnamefont {U.}~\bibnamefont
  {Mukhopadhyay}}, \bibinfo {author} {\bibfnamefont {V.~P.}\ \bibnamefont
  {Michal}}, \bibinfo {author} {\bibfnamefont {Y.}~\bibnamefont {Wang}},
  \bibinfo {author} {\bibfnamefont {B.}~\bibnamefont {Wunsch}}, \bibinfo
  {author} {\bibfnamefont {C.}~\bibnamefont {Reichl}}, \bibinfo {author}
  {\bibfnamefont {W.}~\bibnamefont {Wegscheider}}, \bibinfo {author}
  {\bibfnamefont {M.~S.}\ \bibnamefont {Rudner}}, \bibinfo {author}
  {\bibfnamefont {E.}~\bibnamefont {Demler}}, \ and\ \bibinfo {author}
  {\bibfnamefont {L.~M.~K.}\ \bibnamefont {Vandersypen}},\ }\bibfield  {title}
  {\enquote {\bibinfo {title} {Nagaoka ferromagnetism observed in a quantum dot
  plaquette},}\ }\href {\doibase 10.1038/s41586-020-2051-0} {\bibfield
  {journal} {\bibinfo  {journal} {Nature}\ }\textbf {\bibinfo {volume} {579}},\
  \bibinfo {pages} {528--533} (\bibinfo {year} {2020})}\BibitemShut {NoStop}%
\bibitem [{\citenamefont {Mortemousque}\ \emph {et~al.}(2021)\citenamefont
  {Mortemousque}, \citenamefont {Jadot}, \citenamefont {Chanrion},
  \citenamefont {Thiney}, \citenamefont {Bäuerle}, \citenamefont {Ludwig},
  \citenamefont {Wieck}, \citenamefont {Urdampilleta},\ and\ \citenamefont
  {Meunier}}]{Mortemousque_2021}%
  \BibitemOpen
  \bibfield  {author} {\bibinfo {author} {\bibfnamefont {P.-A.}\ \bibnamefont
  {Mortemousque}}, \bibinfo {author} {\bibfnamefont {B.}~\bibnamefont {Jadot}},
  \bibinfo {author} {\bibfnamefont {E.}~\bibnamefont {Chanrion}}, \bibinfo
  {author} {\bibfnamefont {V.}~\bibnamefont {Thiney}}, \bibinfo {author}
  {\bibfnamefont {C.}~\bibnamefont {Bäuerle}}, \bibinfo {author}
  {\bibfnamefont {A.}~\bibnamefont {Ludwig}}, \bibinfo {author} {\bibfnamefont
  {A.~D.}\ \bibnamefont {Wieck}}, \bibinfo {author} {\bibfnamefont
  {M.}~\bibnamefont {Urdampilleta}}, \ and\ \bibinfo {author} {\bibfnamefont
  {T.}~\bibnamefont {Meunier}},\ }\bibfield  {title} {\enquote {\bibinfo
  {title} {Enhanced spin coherence while displacing electron in a
  two-dimensional array of quantum dots},}\ }\href {\doibase
  10.1103/prxquantum.2.030331} {\bibfield  {journal} {\bibinfo  {journal}
  {{PRX} Quantum}\ }\textbf {\bibinfo {volume} {2}} (\bibinfo {year} {2021}),\
  10.1103/prxquantum.2.030331}\BibitemShut {NoStop}%
\bibitem [{\citenamefont {Fedele}\ \emph {et~al.}(2021)\citenamefont {Fedele},
  \citenamefont {Chatterjee}, \citenamefont {Fallahi}, \citenamefont {Gardner},
  \citenamefont {Manfra},\ and\ \citenamefont {Kuemmeth}}]{Fedele_2021}%
  \BibitemOpen
  \bibfield  {author} {\bibinfo {author} {\bibfnamefont {F.}~\bibnamefont
  {Fedele}}, \bibinfo {author} {\bibfnamefont {A.}~\bibnamefont {Chatterjee}},
  \bibinfo {author} {\bibfnamefont {S.}~\bibnamefont {Fallahi}}, \bibinfo
  {author} {\bibfnamefont {G.~C.}\ \bibnamefont {Gardner}}, \bibinfo {author}
  {\bibfnamefont {M.~J.}\ \bibnamefont {Manfra}}, \ and\ \bibinfo {author}
  {\bibfnamefont {F.}~\bibnamefont {Kuemmeth}},\ }\bibfield  {title} {\enquote
  {\bibinfo {title} {Simultaneous operations in a two-dimensional array of
  singlet-triplet qubits},}\ }\href {\doibase 10.1103/prxquantum.2.040306}
  {\bibfield  {journal} {\bibinfo  {journal} {{PRX} Quantum}\ }\textbf
  {\bibinfo {volume} {2}} (\bibinfo {year} {2021}),\
  10.1103/prxquantum.2.040306}\BibitemShut {NoStop}%
\bibitem [{\citenamefont {Stano}\ and\ \citenamefont {Loss}(2022)}]{Stano2022}%
  \BibitemOpen
  \bibfield  {author} {\bibinfo {author} {\bibfnamefont {P.}~\bibnamefont
  {Stano}}\ and\ \bibinfo {author} {\bibfnamefont {D.}~\bibnamefont {Loss}},\
  }\bibfield  {title} {\enquote {\bibinfo {title} {Review of performance
  metrics of spin qubits in gated semiconducting nanostructures},}\ }\href
  {\doibase 10.1038/s42254-022-00484-w} {\bibfield  {journal} {\bibinfo
  {journal} {Nature Reviews Physics}\ }\textbf {\bibinfo {volume} {4}},\
  \bibinfo {pages} {672--688} (\bibinfo {year} {2022})}\BibitemShut {NoStop}%
\bibitem [{\citenamefont {Hendrickx}\ \emph {et~al.}(2021)\citenamefont
  {Hendrickx}, \citenamefont {Lawrie}, \citenamefont {Russ}, \citenamefont {van
  Riggelen}, \citenamefont {de~Snoo}, \citenamefont {Schouten}, \citenamefont
  {Sammak}, \citenamefont {Scappucci},\ and\ \citenamefont
  {Veldhorst}}]{Hendrickx2021}%
  \BibitemOpen
  \bibfield  {author} {\bibinfo {author} {\bibfnamefont {N.~W.}\ \bibnamefont
  {Hendrickx}}, \bibinfo {author} {\bibfnamefont {W.~I.~L.}\ \bibnamefont
  {Lawrie}}, \bibinfo {author} {\bibfnamefont {M.}~\bibnamefont {Russ}},
  \bibinfo {author} {\bibfnamefont {F.}~\bibnamefont {van Riggelen}}, \bibinfo
  {author} {\bibfnamefont {S.~L.}\ \bibnamefont {de~Snoo}}, \bibinfo {author}
  {\bibfnamefont {R.~N.}\ \bibnamefont {Schouten}}, \bibinfo {author}
  {\bibfnamefont {A.}~\bibnamefont {Sammak}}, \bibinfo {author} {\bibfnamefont
  {G.}~\bibnamefont {Scappucci}}, \ and\ \bibinfo {author} {\bibfnamefont
  {M.}~\bibnamefont {Veldhorst}},\ }\bibfield  {title} {\enquote {\bibinfo
  {title} {A four-qubit germanium quantum processor},}\ }\href {\doibase
  10.1038/s41586-021-03332-6} {\bibfield  {journal} {\bibinfo  {journal}
  {Nature}\ }\textbf {\bibinfo {volume} {591}},\ \bibinfo {pages} {580--585}
  (\bibinfo {year} {2021})}\BibitemShut {NoStop}%
\bibitem [{\citenamefont {van Riggelen}\ \emph {et~al.}(2022)\citenamefont {van
  Riggelen}, \citenamefont {Lawrie}, \citenamefont {Russ}, \citenamefont
  {Hendrickx}, \citenamefont {Sammak}, \citenamefont {Rispler}, \citenamefont
  {Terhal}, \citenamefont {Scappucci},\ and\ \citenamefont
  {Veldhorst}}]{van_Riggelen_2022}%
  \BibitemOpen
  \bibfield  {author} {\bibinfo {author} {\bibfnamefont {F.}~\bibnamefont {van
  Riggelen}}, \bibinfo {author} {\bibfnamefont {W.~I.~L.}\ \bibnamefont
  {Lawrie}}, \bibinfo {author} {\bibfnamefont {M.}~\bibnamefont {Russ}},
  \bibinfo {author} {\bibfnamefont {N.~W.}\ \bibnamefont {Hendrickx}}, \bibinfo
  {author} {\bibfnamefont {A.}~\bibnamefont {Sammak}}, \bibinfo {author}
  {\bibfnamefont {M.}~\bibnamefont {Rispler}}, \bibinfo {author} {\bibfnamefont
  {B.~M.}\ \bibnamefont {Terhal}}, \bibinfo {author} {\bibfnamefont
  {G.}~\bibnamefont {Scappucci}}, \ and\ \bibinfo {author} {\bibfnamefont
  {M.}~\bibnamefont {Veldhorst}},\ }\bibfield  {title} {\enquote {\bibinfo
  {title} {Phase flip code with semiconductor spin qubits},}\ }\href {\doibase
  10.1038/s41534-022-00639-8} {\bibfield  {journal} {\bibinfo  {journal} {npj
  Quantum Information}\ }\textbf {\bibinfo {volume} {8}} (\bibinfo {year}
  {2022}),\ 10.1038/s41534-022-00639-8}\BibitemShut {NoStop}%
\bibitem [{\citenamefont {Borsoi}\ \emph {et~al.}(2022)\citenamefont {Borsoi},
  \citenamefont {Hendrickx}, \citenamefont {John}, \citenamefont {Motz},
  \citenamefont {van Riggelen}, \citenamefont {Sammak}, \citenamefont
  {de~Snoo}, \citenamefont {Scappucci},\ and\ \citenamefont
  {Veldhorst}}]{Borsoi2022}%
  \BibitemOpen
  \bibfield  {author} {\bibinfo {author} {\bibfnamefont {F.}~\bibnamefont
  {Borsoi}}, \bibinfo {author} {\bibfnamefont {N.~W.}\ \bibnamefont
  {Hendrickx}}, \bibinfo {author} {\bibfnamefont {V.}~\bibnamefont {John}},
  \bibinfo {author} {\bibfnamefont {S.}~\bibnamefont {Motz}}, \bibinfo {author}
  {\bibfnamefont {F.}~\bibnamefont {van Riggelen}}, \bibinfo {author}
  {\bibfnamefont {A.}~\bibnamefont {Sammak}}, \bibinfo {author} {\bibfnamefont
  {S.~L.}\ \bibnamefont {de~Snoo}}, \bibinfo {author} {\bibfnamefont
  {G.}~\bibnamefont {Scappucci}}, \ and\ \bibinfo {author} {\bibfnamefont
  {M.}~\bibnamefont {Veldhorst}},\ }\bibfield  {title} {\enquote {\bibinfo
  {title} {Shared control of a 16 semiconductor quantum dot crossbar array},}\
  }\href {\doibase 10.48550/ARXIV.2209.06609} {\bibfield  {journal} {\bibinfo
  {journal} {arXiv}\ } (\bibinfo {year} {2022}),\
  10.48550/ARXIV.2209.06609}\BibitemShut {NoStop}%
\bibitem [{\citenamefont {Gilbert}\ \emph {et~al.}(2020)\citenamefont
  {Gilbert}, \citenamefont {Saraiva}, \citenamefont {Lim}, \citenamefont
  {Yang}, \citenamefont {Laucht}, \citenamefont {Bertrand}, \citenamefont
  {Rambal}, \citenamefont {Hutin}, \citenamefont {Escott}, \citenamefont
  {Vinet},\ and\ \citenamefont {Dzurak}}]{Gilbert_2020}%
  \BibitemOpen
  \bibfield  {author} {\bibinfo {author} {\bibfnamefont {W.}~\bibnamefont
  {Gilbert}}, \bibinfo {author} {\bibfnamefont {A.}~\bibnamefont {Saraiva}},
  \bibinfo {author} {\bibfnamefont {W.~H.}\ \bibnamefont {Lim}}, \bibinfo
  {author} {\bibfnamefont {C.~H.}\ \bibnamefont {Yang}}, \bibinfo {author}
  {\bibfnamefont {A.}~\bibnamefont {Laucht}}, \bibinfo {author} {\bibfnamefont
  {B.}~\bibnamefont {Bertrand}}, \bibinfo {author} {\bibfnamefont
  {N.}~\bibnamefont {Rambal}}, \bibinfo {author} {\bibfnamefont
  {L.}~\bibnamefont {Hutin}}, \bibinfo {author} {\bibfnamefont {C.~C.}\
  \bibnamefont {Escott}}, \bibinfo {author} {\bibfnamefont {M.}~\bibnamefont
  {Vinet}}, \ and\ \bibinfo {author} {\bibfnamefont {A.~S.}\ \bibnamefont
  {Dzurak}},\ }\bibfield  {title} {\enquote {\bibinfo {title} {Single-electron
  operation of a silicon-{CMOS} 2 {\texttimes} 2 quantum dot array with
  integrated charge sensing},}\ }\href {\doibase 10.1021/acs.nanolett.0c02397}
  {\bibfield  {journal} {\bibinfo  {journal} {Nano Letters}\ }\textbf {\bibinfo
  {volume} {20}},\ \bibinfo {pages} {7882--7888} (\bibinfo {year}
  {2020})}\BibitemShut {NoStop}%
\bibitem [{\citenamefont {Chanrion}\ \emph {et~al.}(2020)\citenamefont
  {Chanrion}, \citenamefont {Niegemann}, \citenamefont {Bertrand},
  \citenamefont {Spence}, \citenamefont {Jadot}, \citenamefont {Li},
  \citenamefont {Mortemousque}, \citenamefont {Hutin}, \citenamefont {Maurand},
  \citenamefont {Jehl}, \citenamefont {Sanquer}, \citenamefont {Franceschi},
  \citenamefont {Bäuerle}, \citenamefont {Balestro}, \citenamefont {Niquet},
  \citenamefont {Vinet}, \citenamefont {Meunier},\ and\ \citenamefont
  {Urdampilleta}}]{Chanrion_2020}%
  \BibitemOpen
  \bibfield  {author} {\bibinfo {author} {\bibfnamefont {E.}~\bibnamefont
  {Chanrion}}, \bibinfo {author} {\bibfnamefont {D.~J.}\ \bibnamefont
  {Niegemann}}, \bibinfo {author} {\bibfnamefont {B.}~\bibnamefont {Bertrand}},
  \bibinfo {author} {\bibfnamefont {C.}~\bibnamefont {Spence}}, \bibinfo
  {author} {\bibfnamefont {B.}~\bibnamefont {Jadot}}, \bibinfo {author}
  {\bibfnamefont {J.}~\bibnamefont {Li}}, \bibinfo {author} {\bibfnamefont
  {P.-A.}\ \bibnamefont {Mortemousque}}, \bibinfo {author} {\bibfnamefont
  {L.}~\bibnamefont {Hutin}}, \bibinfo {author} {\bibfnamefont
  {R.}~\bibnamefont {Maurand}}, \bibinfo {author} {\bibfnamefont
  {X.}~\bibnamefont {Jehl}}, \bibinfo {author} {\bibfnamefont {M.}~\bibnamefont
  {Sanquer}}, \bibinfo {author} {\bibfnamefont {S.~D.}\ \bibnamefont
  {Franceschi}}, \bibinfo {author} {\bibfnamefont {C.}~\bibnamefont
  {Bäuerle}}, \bibinfo {author} {\bibfnamefont {F.}~\bibnamefont {Balestro}},
  \bibinfo {author} {\bibfnamefont {Y.-M.}\ \bibnamefont {Niquet}}, \bibinfo
  {author} {\bibfnamefont {M.}~\bibnamefont {Vinet}}, \bibinfo {author}
  {\bibfnamefont {T.}~\bibnamefont {Meunier}}, \ and\ \bibinfo {author}
  {\bibfnamefont {M.}~\bibnamefont {Urdampilleta}},\ }\bibfield  {title}
  {\enquote {\bibinfo {title} {Charge detection in an array of {CMOS} quantum
  dots},}\ }\href {\doibase 10.1103/physrevapplied.14.024066} {\bibfield
  {journal} {\bibinfo  {journal} {Physical Review Applied}\ }\textbf {\bibinfo
  {volume} {14}} (\bibinfo {year} {2020}),\
  10.1103/physrevapplied.14.024066}\BibitemShut {NoStop}%
\bibitem [{\citenamefont {Martins}\ \emph {et~al.}(2016)\citenamefont
  {Martins}, \citenamefont {Malinowski}, \citenamefont {Nissen}, \citenamefont
  {Barnes}, \citenamefont {Fallahi}, \citenamefont {Gardner}, \citenamefont
  {Manfra}, \citenamefont {Marcus},\ and\ \citenamefont
  {Kuemmeth}}]{Martins2016}%
  \BibitemOpen
  \bibfield  {author} {\bibinfo {author} {\bibfnamefont {F.}~\bibnamefont
  {Martins}}, \bibinfo {author} {\bibfnamefont {F.~K.}\ \bibnamefont
  {Malinowski}}, \bibinfo {author} {\bibfnamefont {P.~D.}\ \bibnamefont
  {Nissen}}, \bibinfo {author} {\bibfnamefont {E.}~\bibnamefont {Barnes}},
  \bibinfo {author} {\bibfnamefont {S.}~\bibnamefont {Fallahi}}, \bibinfo
  {author} {\bibfnamefont {G.~C.}\ \bibnamefont {Gardner}}, \bibinfo {author}
  {\bibfnamefont {M.~J.}\ \bibnamefont {Manfra}}, \bibinfo {author}
  {\bibfnamefont {C.~M.}\ \bibnamefont {Marcus}}, \ and\ \bibinfo {author}
  {\bibfnamefont {F.}~\bibnamefont {Kuemmeth}},\ }\bibfield  {title} {\enquote
  {\bibinfo {title} {Noise suppression using symmetric exchange gates in spin
  qubits},}\ }\href {\doibase 10.1103/physrevlett.116.116801} {\bibfield
  {journal} {\bibinfo  {journal} {Physical Review Letters}\ }\textbf {\bibinfo
  {volume} {116}} (\bibinfo {year} {2016}),\
  10.1103/physrevlett.116.116801}\BibitemShut {NoStop}%
\bibitem [{\citenamefont {Reed}\ \emph {et~al.}(2016)\citenamefont {Reed},
  \citenamefont {Maune}, \citenamefont {Andrews}, \citenamefont {Borselli},
  \citenamefont {Eng}, \citenamefont {Jura}, \citenamefont {Kiselev},
  \citenamefont {Ladd}, \citenamefont {Merkel}, \citenamefont {Milosavljevic},
  \citenamefont {Pritchett}, \citenamefont {Rakher}, \citenamefont {Ross},
  \citenamefont {Schmitz}, \citenamefont {Smith}, \citenamefont {Wright},
  \citenamefont {Gyure},\ and\ \citenamefont {Hunter}}]{Reed2016}%
  \BibitemOpen
  \bibfield  {author} {\bibinfo {author} {\bibfnamefont {M.}~\bibnamefont
  {Reed}}, \bibinfo {author} {\bibfnamefont {B.}~\bibnamefont {Maune}},
  \bibinfo {author} {\bibfnamefont {R.}~\bibnamefont {Andrews}}, \bibinfo
  {author} {\bibfnamefont {M.}~\bibnamefont {Borselli}}, \bibinfo {author}
  {\bibfnamefont {K.}~\bibnamefont {Eng}}, \bibinfo {author} {\bibfnamefont
  {M.}~\bibnamefont {Jura}}, \bibinfo {author} {\bibfnamefont {A.}~\bibnamefont
  {Kiselev}}, \bibinfo {author} {\bibfnamefont {T.}~\bibnamefont {Ladd}},
  \bibinfo {author} {\bibfnamefont {S.}~\bibnamefont {Merkel}}, \bibinfo
  {author} {\bibfnamefont {I.}~\bibnamefont {Milosavljevic}}, \bibinfo {author}
  {\bibfnamefont {E.}~\bibnamefont {Pritchett}}, \bibinfo {author}
  {\bibfnamefont {M.}~\bibnamefont {Rakher}}, \bibinfo {author} {\bibfnamefont
  {R.}~\bibnamefont {Ross}}, \bibinfo {author} {\bibfnamefont {A.}~\bibnamefont
  {Schmitz}}, \bibinfo {author} {\bibfnamefont {A.}~\bibnamefont {Smith}},
  \bibinfo {author} {\bibfnamefont {J.}~\bibnamefont {Wright}}, \bibinfo
  {author} {\bibfnamefont {M.}~\bibnamefont {Gyure}}, \ and\ \bibinfo {author}
  {\bibfnamefont {A.}~\bibnamefont {Hunter}},\ }\bibfield  {title} {\enquote
  {\bibinfo {title} {Reduced sensitivity to charge noise in semiconductor spin
  qubits via symmetric operation},}\ }\href {\doibase
  10.1103/physrevlett.116.110402} {\bibfield  {journal} {\bibinfo  {journal}
  {Physical Review Letters}\ }\textbf {\bibinfo {volume} {116}} (\bibinfo
  {year} {2016}),\ 10.1103/physrevlett.116.110402}\BibitemShut {NoStop}%
\bibitem [{\citenamefont {Lawrie}\ \emph {et~al.}(2020)\citenamefont {Lawrie},
  \citenamefont {Eenink}, \citenamefont {Hendrickx}, \citenamefont {Boter},
  \citenamefont {Petit}, \citenamefont {Amitonov}, \citenamefont {Lodari},
  \citenamefont {Wuetz}, \citenamefont {Volk}, \citenamefont {Philips},
  \citenamefont {Droulers}, \citenamefont {Kalhor}, \citenamefont {van
  Riggelen}, \citenamefont {Brousse}, \citenamefont {Sammak}, \citenamefont
  {Vandersypen}, \citenamefont {Scappucci},\ and\ \citenamefont
  {Veldhorst}}]{Lawrie2020}%
  \BibitemOpen
  \bibfield  {author} {\bibinfo {author} {\bibfnamefont {W.~I.~L.}\
  \bibnamefont {Lawrie}}, \bibinfo {author} {\bibfnamefont {H.~G.~J.}\
  \bibnamefont {Eenink}}, \bibinfo {author} {\bibfnamefont {N.~W.}\
  \bibnamefont {Hendrickx}}, \bibinfo {author} {\bibfnamefont {J.~M.}\
  \bibnamefont {Boter}}, \bibinfo {author} {\bibfnamefont {L.}~\bibnamefont
  {Petit}}, \bibinfo {author} {\bibfnamefont {S.~V.}\ \bibnamefont {Amitonov}},
  \bibinfo {author} {\bibfnamefont {M.}~\bibnamefont {Lodari}}, \bibinfo
  {author} {\bibfnamefont {B.~P.}\ \bibnamefont {Wuetz}}, \bibinfo {author}
  {\bibfnamefont {C.}~\bibnamefont {Volk}}, \bibinfo {author} {\bibfnamefont
  {S.~G.~J.}\ \bibnamefont {Philips}}, \bibinfo {author} {\bibfnamefont
  {G.}~\bibnamefont {Droulers}}, \bibinfo {author} {\bibfnamefont
  {N.}~\bibnamefont {Kalhor}}, \bibinfo {author} {\bibfnamefont
  {F.}~\bibnamefont {van Riggelen}}, \bibinfo {author} {\bibfnamefont
  {D.}~\bibnamefont {Brousse}}, \bibinfo {author} {\bibfnamefont
  {A.}~\bibnamefont {Sammak}}, \bibinfo {author} {\bibfnamefont {L.~M.~K.}\
  \bibnamefont {Vandersypen}}, \bibinfo {author} {\bibfnamefont
  {G.}~\bibnamefont {Scappucci}}, \ and\ \bibinfo {author} {\bibfnamefont
  {M.}~\bibnamefont {Veldhorst}},\ }\bibfield  {title} {\enquote {\bibinfo
  {title} {Quantum dot arrays in silicon and germanium},}\ }\href {\doibase
  10.1063/5.0002013} {\bibfield  {journal} {\bibinfo  {journal} {Applied
  Physics Letters}\ }\textbf {\bibinfo {volume} {116}},\ \bibinfo {pages}
  {080501} (\bibinfo {year} {2020})}\BibitemShut {NoStop}%
\bibitem [{\citenamefont {Volk}\ \emph {et~al.}(2019)\citenamefont {Volk},
  \citenamefont {Zwerver}, \citenamefont {Mukhopadhyay}, \citenamefont
  {Eendebak}, \citenamefont {van Diepen}, \citenamefont {Dehollain},
  \citenamefont {Hensgens}, \citenamefont {Fujita}, \citenamefont {Reichl},
  \citenamefont {Wegscheider},\ and\ \citenamefont {Vandersypen}}]{Volk2019}%
  \BibitemOpen
  \bibfield  {author} {\bibinfo {author} {\bibfnamefont {C.}~\bibnamefont
  {Volk}}, \bibinfo {author} {\bibfnamefont {A.~M.~J.}\ \bibnamefont
  {Zwerver}}, \bibinfo {author} {\bibfnamefont {U.}~\bibnamefont
  {Mukhopadhyay}}, \bibinfo {author} {\bibfnamefont {P.~T.}\ \bibnamefont
  {Eendebak}}, \bibinfo {author} {\bibfnamefont {C.~J.}\ \bibnamefont {van
  Diepen}}, \bibinfo {author} {\bibfnamefont {J.~P.}\ \bibnamefont
  {Dehollain}}, \bibinfo {author} {\bibfnamefont {T.}~\bibnamefont {Hensgens}},
  \bibinfo {author} {\bibfnamefont {T.}~\bibnamefont {Fujita}}, \bibinfo
  {author} {\bibfnamefont {C.}~\bibnamefont {Reichl}}, \bibinfo {author}
  {\bibfnamefont {W.}~\bibnamefont {Wegscheider}}, \ and\ \bibinfo {author}
  {\bibfnamefont {L.~M.~K.}\ \bibnamefont {Vandersypen}},\ }\bibfield  {title}
  {\enquote {\bibinfo {title} {Loading a quantum-dot based
  {\textquotedblleft}qubyte{\textquotedblright} register},}\ }\href {\doibase
  10.1038/s41534-019-0146-y} {\bibfield  {journal} {\bibinfo  {journal} {npj
  Quantum Information}\ }\textbf {\bibinfo {volume} {5}} (\bibinfo {year}
  {2019}),\ 10.1038/s41534-019-0146-y}\BibitemShut {NoStop}%
\bibitem [{\citenamefont {DiCarlo}\ \emph {et~al.}(2004)\citenamefont
  {DiCarlo}, \citenamefont {Lynch}, \citenamefont {Johnson}, \citenamefont
  {Childress}, \citenamefont {Crockett}, \citenamefont {Marcus}, \citenamefont
  {Hanson},\ and\ \citenamefont {Gossard}}]{DiCarlo2004}%
  \BibitemOpen
  \bibfield  {author} {\bibinfo {author} {\bibfnamefont {L.}~\bibnamefont
  {DiCarlo}}, \bibinfo {author} {\bibfnamefont {H.~J.}\ \bibnamefont {Lynch}},
  \bibinfo {author} {\bibfnamefont {A.~C.}\ \bibnamefont {Johnson}}, \bibinfo
  {author} {\bibfnamefont {L.~I.}\ \bibnamefont {Childress}}, \bibinfo {author}
  {\bibfnamefont {K.}~\bibnamefont {Crockett}}, \bibinfo {author}
  {\bibfnamefont {C.~M.}\ \bibnamefont {Marcus}}, \bibinfo {author}
  {\bibfnamefont {M.~P.}\ \bibnamefont {Hanson}}, \ and\ \bibinfo {author}
  {\bibfnamefont {A.~C.}\ \bibnamefont {Gossard}},\ }\bibfield  {title}
  {\enquote {\bibinfo {title} {Differential charge sensing and charge
  delocalization in a tunable double quantum dot},}\ }\href {\doibase
  10.1103/physrevlett.92.226801} {\bibfield  {journal} {\bibinfo  {journal}
  {Physical Review Letters}\ }\textbf {\bibinfo {volume} {92}} (\bibinfo {year}
  {2004}),\ 10.1103/physrevlett.92.226801}\BibitemShut {NoStop}%
\bibitem [{\citenamefont {van Diepen}\ \emph {et~al.}(2018)\citenamefont {van
  Diepen}, \citenamefont {Eendebak}, \citenamefont {Buijtendorp}, \citenamefont
  {Mukhopadhyay}, \citenamefont {Fujita}, \citenamefont {Reichl}, \citenamefont
  {Wegscheider},\ and\ \citenamefont {Vandersypen}}]{vanDiepen2018}%
  \BibitemOpen
  \bibfield  {author} {\bibinfo {author} {\bibfnamefont {C.~J.}\ \bibnamefont
  {van Diepen}}, \bibinfo {author} {\bibfnamefont {P.~T.}\ \bibnamefont
  {Eendebak}}, \bibinfo {author} {\bibfnamefont {B.~T.}\ \bibnamefont
  {Buijtendorp}}, \bibinfo {author} {\bibfnamefont {U.}~\bibnamefont
  {Mukhopadhyay}}, \bibinfo {author} {\bibfnamefont {T.}~\bibnamefont
  {Fujita}}, \bibinfo {author} {\bibfnamefont {C.}~\bibnamefont {Reichl}},
  \bibinfo {author} {\bibfnamefont {W.}~\bibnamefont {Wegscheider}}, \ and\
  \bibinfo {author} {\bibfnamefont {L.~M.~K.}\ \bibnamefont {Vandersypen}},\
  }\bibfield  {title} {\enquote {\bibinfo {title} {Automated tuning of
  inter-dot tunnel coupling in double quantum dots},}\ }\href {\doibase
  10.1063/1.5031034} {\bibfield  {journal} {\bibinfo  {journal} {Applied
  Physics Letters}\ }\textbf {\bibinfo {volume} {113}},\ \bibinfo {pages}
  {033101} (\bibinfo {year} {2018})}\BibitemShut {NoStop}%
\bibitem [{\citenamefont {Knapp}\ \emph {et~al.}(2016)\citenamefont {Knapp},
  \citenamefont {Mohr}, \citenamefont {Li}, \citenamefont {Thorgrimsson},
  \citenamefont {Foote}, \citenamefont {Wu}, \citenamefont {Ward},
  \citenamefont {Savage}, \citenamefont {Lagally}, \citenamefont {Friesen},
  \citenamefont {Coppersmith},\ and\ \citenamefont {Eriksson}}]{Knapp2016}%
  \BibitemOpen
  \bibfield  {author} {\bibinfo {author} {\bibfnamefont {T.~J.}\ \bibnamefont
  {Knapp}}, \bibinfo {author} {\bibfnamefont {R.~T.}\ \bibnamefont {Mohr}},
  \bibinfo {author} {\bibfnamefont {Y.~S.}\ \bibnamefont {Li}}, \bibinfo
  {author} {\bibfnamefont {B.}~\bibnamefont {Thorgrimsson}}, \bibinfo {author}
  {\bibfnamefont {R.~H.}\ \bibnamefont {Foote}}, \bibinfo {author}
  {\bibfnamefont {X.}~\bibnamefont {Wu}}, \bibinfo {author} {\bibfnamefont
  {D.~R.}\ \bibnamefont {Ward}}, \bibinfo {author} {\bibfnamefont {D.~E.}\
  \bibnamefont {Savage}}, \bibinfo {author} {\bibfnamefont {M.~G.}\
  \bibnamefont {Lagally}}, \bibinfo {author} {\bibfnamefont {M.}~\bibnamefont
  {Friesen}}, \bibinfo {author} {\bibfnamefont {S.~N.}\ \bibnamefont
  {Coppersmith}}, \ and\ \bibinfo {author} {\bibfnamefont {M.~A.}\ \bibnamefont
  {Eriksson}},\ }\bibfield  {title} {\enquote {\bibinfo {title}
  {Characterization of a gate-defined double quantum dot in a {Si} /{SiGe}
  nanomembrane},}\ }\href {\doibase 10.1088/0957-4484/27/15/154002} {\bibfield
  {journal} {\bibinfo  {journal} {Nanotechnology}\ }\textbf {\bibinfo {volume}
  {27}},\ \bibinfo {pages} {154002} (\bibinfo {year} {2016})}\BibitemShut
  {NoStop}%
\bibitem [{\citenamefont {Mukhopadhyay}\ \emph {et~al.}(2018)\citenamefont
  {Mukhopadhyay}, \citenamefont {Dehollain}, \citenamefont {Reichl},
  \citenamefont {Wegscheider},\ and\ \citenamefont
  {Vandersypen}}]{Mukhopadhyay2018}%
  \BibitemOpen
  \bibfield  {author} {\bibinfo {author} {\bibfnamefont {U.}~\bibnamefont
  {Mukhopadhyay}}, \bibinfo {author} {\bibfnamefont {J.~P.}\ \bibnamefont
  {Dehollain}}, \bibinfo {author} {\bibfnamefont {C.}~\bibnamefont {Reichl}},
  \bibinfo {author} {\bibfnamefont {W.}~\bibnamefont {Wegscheider}}, \ and\
  \bibinfo {author} {\bibfnamefont {L.~M.~K.}\ \bibnamefont {Vandersypen}},\
  }\bibfield  {title} {\enquote {\bibinfo {title} {A
  2{\hspace{0.167em}}{\texttimes}{\hspace{0.167em}}2 quantum dot array with
  controllable inter-dot tunnel couplings},}\ }\href {\doibase
  10.1063/1.5025928} {\bibfield  {journal} {\bibinfo  {journal} {Applied
  Physics Letters}\ }\textbf {\bibinfo {volume} {112}},\ \bibinfo {pages}
  {183505} (\bibinfo {year} {2018})}\BibitemShut {NoStop}%
\bibitem [{\citenamefont {Ha}\ \emph {et~al.}(2021)\citenamefont {Ha},
  \citenamefont {Ha}, \citenamefont {Choi}, \citenamefont {Tang}, \citenamefont
  {Schmitz}, \citenamefont {Levendorf}, \citenamefont {Lee}, \citenamefont
  {Chappell}, \citenamefont {Adams}, \citenamefont {Hulbert}, \citenamefont
  {Acuna}, \citenamefont {Noah}, \citenamefont {Matten}, \citenamefont {Jura},
  \citenamefont {Wright}, \citenamefont {Rakher},\ and\ \citenamefont
  {Borselli}}]{Ha2021}%
  \BibitemOpen
  \bibfield  {author} {\bibinfo {author} {\bibfnamefont {W.}~\bibnamefont
  {Ha}}, \bibinfo {author} {\bibfnamefont {S.~D.}\ \bibnamefont {Ha}}, \bibinfo
  {author} {\bibfnamefont {M.~D.}\ \bibnamefont {Choi}}, \bibinfo {author}
  {\bibfnamefont {Y.}~\bibnamefont {Tang}}, \bibinfo {author} {\bibfnamefont
  {A.~E.}\ \bibnamefont {Schmitz}}, \bibinfo {author} {\bibfnamefont {M.~P.}\
  \bibnamefont {Levendorf}}, \bibinfo {author} {\bibfnamefont {K.}~\bibnamefont
  {Lee}}, \bibinfo {author} {\bibfnamefont {J.~M.}\ \bibnamefont {Chappell}},
  \bibinfo {author} {\bibfnamefont {T.~S.}\ \bibnamefont {Adams}}, \bibinfo
  {author} {\bibfnamefont {D.~R.}\ \bibnamefont {Hulbert}}, \bibinfo {author}
  {\bibfnamefont {E.}~\bibnamefont {Acuna}}, \bibinfo {author} {\bibfnamefont
  {R.~S.}\ \bibnamefont {Noah}}, \bibinfo {author} {\bibfnamefont {J.~W.}\
  \bibnamefont {Matten}}, \bibinfo {author} {\bibfnamefont {M.~P.}\
  \bibnamefont {Jura}}, \bibinfo {author} {\bibfnamefont {J.~A.}\ \bibnamefont
  {Wright}}, \bibinfo {author} {\bibfnamefont {M.~T.}\ \bibnamefont {Rakher}},
  \ and\ \bibinfo {author} {\bibfnamefont {M.~G.}\ \bibnamefont {Borselli}},\
  }\bibfield  {title} {\enquote {\bibinfo {title} {A flexible design platform
  for si/{SiGe} exchange-only qubits with low disorder},}\ }\href {\doibase
  10.1021/acs.nanolett.1c03026} {\bibfield  {journal} {\bibinfo  {journal}
  {Nano Letters}\ }\textbf {\bibinfo {volume} {22}},\ \bibinfo {pages}
  {1443--1448} (\bibinfo {year} {2021})}\BibitemShut {NoStop}%
\bibitem [{\citenamefont {Veldhorst}\ \emph {et~al.}(2017)\citenamefont
  {Veldhorst}, \citenamefont {Eenink}, \citenamefont {Yang},\ and\
  \citenamefont {Dzurak}}]{Veldhorst2017}%
  \BibitemOpen
  \bibfield  {author} {\bibinfo {author} {\bibfnamefont {M.}~\bibnamefont
  {Veldhorst}}, \bibinfo {author} {\bibfnamefont {H.~G.~J.}\ \bibnamefont
  {Eenink}}, \bibinfo {author} {\bibfnamefont {C.~H.}\ \bibnamefont {Yang}}, \
  and\ \bibinfo {author} {\bibfnamefont {A.~S.}\ \bibnamefont {Dzurak}},\
  }\bibfield  {title} {\enquote {\bibinfo {title} {Silicon {CMOS} architecture
  for a spin-based quantum computer},}\ }\href {\doibase
  10.1038/s41467-017-01905-6} {\bibfield  {journal} {\bibinfo  {journal}
  {Nature Communications}\ }\textbf {\bibinfo {volume} {8}} (\bibinfo {year}
  {2017}),\ 10.1038/s41467-017-01905-6}\BibitemShut {NoStop}%
\bibitem [{\citenamefont {Li}\ \emph {et~al.}(2018)\citenamefont {Li},
  \citenamefont {Petit}, \citenamefont {Franke}, \citenamefont {Dehollain},
  \citenamefont {Helsen}, \citenamefont {Steudtner}, \citenamefont {Thomas},
  \citenamefont {Yoscovits}, \citenamefont {Singh}, \citenamefont {Wehner},
  \citenamefont {Vandersypen}, \citenamefont {Clarke},\ and\ \citenamefont
  {Veldhorst}}]{Li2018}%
  \BibitemOpen
  \bibfield  {author} {\bibinfo {author} {\bibfnamefont {R.}~\bibnamefont
  {Li}}, \bibinfo {author} {\bibfnamefont {L.}~\bibnamefont {Petit}}, \bibinfo
  {author} {\bibfnamefont {D.~P.}\ \bibnamefont {Franke}}, \bibinfo {author}
  {\bibfnamefont {J.~P.}\ \bibnamefont {Dehollain}}, \bibinfo {author}
  {\bibfnamefont {J.}~\bibnamefont {Helsen}}, \bibinfo {author} {\bibfnamefont
  {M.}~\bibnamefont {Steudtner}}, \bibinfo {author} {\bibfnamefont {N.~K.}\
  \bibnamefont {Thomas}}, \bibinfo {author} {\bibfnamefont {Z.~R.}\
  \bibnamefont {Yoscovits}}, \bibinfo {author} {\bibfnamefont {K.~J.}\
  \bibnamefont {Singh}}, \bibinfo {author} {\bibfnamefont {S.}~\bibnamefont
  {Wehner}}, \bibinfo {author} {\bibfnamefont {L.~M.~K.}\ \bibnamefont
  {Vandersypen}}, \bibinfo {author} {\bibfnamefont {J.~S.}\ \bibnamefont
  {Clarke}}, \ and\ \bibinfo {author} {\bibfnamefont {M.}~\bibnamefont
  {Veldhorst}},\ }\bibfield  {title} {\enquote {\bibinfo {title} {A crossbar
  network for silicon quantum dot qubits},}\ }\href {\doibase
  10.1126/sciadv.aar3960} {\bibfield  {journal} {\bibinfo  {journal} {Science
  Advances}\ }\textbf {\bibinfo {volume} {4}} (\bibinfo {year} {2018}),\
  10.1126/sciadv.aar3960}\BibitemShut {NoStop}%
\bibitem [{\citenamefont {Unseld}\ \emph {et~al.}(2023)\citenamefont {Unseld},
  \citenamefont {Meyer}, \citenamefont {Madzik}, \citenamefont {Borsoi},
  \citenamefont {de~Snoo}, \citenamefont {Amitonov}, \citenamefont {Sammak},
  \citenamefont {Scappucci}, \citenamefont {Veldhorst},\ and\ \citenamefont
  {Vandersypen}}]{Dataset}%
  \BibitemOpen
  \bibfield  {author} {\bibinfo {author} {\bibfnamefont {F.~K.}\ \bibnamefont
  {Unseld}}, \bibinfo {author} {\bibfnamefont {M.}~\bibnamefont {Meyer}},
  \bibinfo {author} {\bibfnamefont {M.}~\bibnamefont {Madzik}}, \bibinfo
  {author} {\bibfnamefont {F.}~\bibnamefont {Borsoi}}, \bibinfo {author}
  {\bibfnamefont {S.}~\bibnamefont {de~Snoo}}, \bibinfo {author} {\bibfnamefont
  {S.}~\bibnamefont {Amitonov}}, \bibinfo {author} {\bibfnamefont
  {A.}~\bibnamefont {Sammak}}, \bibinfo {author} {\bibfnamefont
  {G.}~\bibnamefont {Scappucci}}, \bibinfo {author} {\bibfnamefont
  {M.}~\bibnamefont {Veldhorst}}, \ and\ \bibinfo {author} {\bibfnamefont
  {L.}~\bibnamefont {Vandersypen}},\ }\href
  {https://doi.org/10.5281/zenodo.7957631} {\enquote {\bibinfo {title}
  {{Dataset underlying the manuscript: A 2D quantum dot array in planar
  28Si/{SiGe}}},}\ }\bibinfo {howpublished} {Zenodo} (\bibinfo {year} {2023}),\
  \bibinfo {note} {10.5281/zenodo.7957631}\BibitemShut {NoStop}%
\bibitem [{\citenamefont {Esposti}\ \emph {et~al.}(2022)\citenamefont
  {Esposti}, \citenamefont {Wuetz}, \citenamefont {Fezzi}, \citenamefont
  {Lodari}, \citenamefont {Sammak},\ and\ \citenamefont
  {Scappucci}}]{DegliEsposti2022}%
  \BibitemOpen
  \bibfield  {author} {\bibinfo {author} {\bibfnamefont {D.~D.}\ \bibnamefont
  {Esposti}}, \bibinfo {author} {\bibfnamefont {B.~P.}\ \bibnamefont {Wuetz}},
  \bibinfo {author} {\bibfnamefont {V.}~\bibnamefont {Fezzi}}, \bibinfo
  {author} {\bibfnamefont {M.}~\bibnamefont {Lodari}}, \bibinfo {author}
  {\bibfnamefont {A.}~\bibnamefont {Sammak}}, \ and\ \bibinfo {author}
  {\bibfnamefont {G.}~\bibnamefont {Scappucci}},\ }\bibfield  {title} {\enquote
  {\bibinfo {title} {Wafer-scale low-disorder 2deg in 28si/{SiGe} without an
  epitaxial si cap},}\ }\href {\doibase 10.1063/5.0088576} {\bibfield
  {journal} {\bibinfo  {journal} {Applied Physics Letters}\ }\textbf {\bibinfo
  {volume} {120}} (\bibinfo {year} {2022}),\ 10.1063/5.0088576}\BibitemShut
  {NoStop}%
\bibitem [{\citenamefont {Petit}\ \emph {et~al.}(2018)\citenamefont {Petit},
  \citenamefont {Boter}, \citenamefont {Eenink}, \citenamefont {Droulers},
  \citenamefont {Tagliaferri}, \citenamefont {Li}, \citenamefont {Franke},
  \citenamefont {Singh}, \citenamefont {Clarke}, \citenamefont {Schouten},
  \citenamefont {Dobrovitski}, \citenamefont {Vandersypen},\ and\ \citenamefont
  {Veldhorst}}]{Petit2018}%
  \BibitemOpen
  \bibfield  {author} {\bibinfo {author} {\bibfnamefont {L.}~\bibnamefont
  {Petit}}, \bibinfo {author} {\bibfnamefont {J.}~\bibnamefont {Boter}},
  \bibinfo {author} {\bibfnamefont {H.}~\bibnamefont {Eenink}}, \bibinfo
  {author} {\bibfnamefont {G.}~\bibnamefont {Droulers}}, \bibinfo {author}
  {\bibfnamefont {M.}~\bibnamefont {Tagliaferri}}, \bibinfo {author}
  {\bibfnamefont {R.}~\bibnamefont {Li}}, \bibinfo {author} {\bibfnamefont
  {D.}~\bibnamefont {Franke}}, \bibinfo {author} {\bibfnamefont
  {K.}~\bibnamefont {Singh}}, \bibinfo {author} {\bibfnamefont
  {J.}~\bibnamefont {Clarke}}, \bibinfo {author} {\bibfnamefont
  {R.}~\bibnamefont {Schouten}}, \bibinfo {author} {\bibfnamefont
  {V.}~\bibnamefont {Dobrovitski}}, \bibinfo {author} {\bibfnamefont
  {L.}~\bibnamefont {Vandersypen}}, \ and\ \bibinfo {author} {\bibfnamefont
  {M.}~\bibnamefont {Veldhorst}},\ }\bibfield  {title} {\enquote {\bibinfo
  {title} {Spin lifetime and charge noise in hot silicon quantum dot qubits},}\
  }\href {\doibase 10.1103/physrevlett.121.076801} {\bibfield  {journal}
  {\bibinfo  {journal} {Physical Review Letters}\ }\textbf {\bibinfo {volume}
  {121}} (\bibinfo {year} {2018}),\ 10.1103/physrevlett.121.076801}\BibitemShut
  {NoStop}%
\end{thebibliography}%

\end{document}